\documentstyle[12pt,aaspp4,flushrt,graphicx]{article}

\def\etal{et~al.\,}
\def\Msun{${\rm M}_\odot$}                  

\def\drp{\Delta m_{15}(B)}

\def\Hunits {km s$^{-1}$ Mpc$^{-1}$}
\def\kps {km s$^{-1}$}

\def\plotone#1{\centering \leavevmode
\epsfxsize=\textwidth \epsfbox{#1}}

\def\hof{H\"oflich}

\def\eg{e.\,g.\,}
\def\ie{i.\,e.\,}
\def\Ni56{$^{56}$Ni}
\def\Co56{$^{56}$Co}
\def\M56Ni{$M_{^{56}\rm Ni}$}
\def\OI{\ion{O}{1}}
\def\SiII{\ion{Si}{2}}
\def\SII{\ion{S}{2}}
\def\MgII{\ion{Mg}{2}}

\def\FeII{\ion{Fe}{2}}
\def\TiII{\ion{Ti}{2}}
\def\MgII{\ion{Mg}{2}}

\def\qu{$Q-U$}

\def\deg{\hbox{$^\circ$}}

\lefthead{Howell, H\"oflich, Wang, \& Wheeler}
\righthead{SN 1999by}

\begin{document}

\begin{center}
\vspace{3.5mm}
\title{Evidence for Asphericity in a Subluminous Type Ia Supernova: 
Spectropolarimetry of SN 1999by}
\author{D. Andrew Howell\altaffilmark{1}, Peter H\"oflich, Lifan Wang\altaffilmark{1}, and J. Craig Wheeler} 
\affil{Department of Astronomy and McDonald Observatory, University of Texas at Austin, Austin, TX 78712\\
howell, pah, lifan, wheel@astro.as.utexas.edu}
\authoraddr{Astronomy Department, University of Texas, RLM 15.308, Austin, TX 78712-1083} 
\altaffiltext{1}{Present address: Institute for Nuclear and Particle 
Astrophysics, E. O. Lawrence Berkeley National Laboratory, Berkeley CA 94720}

\end{center}

\begin{abstract}

 We present polarization spectra near maximum light
for the strongly subluminous Type Ia supernova 1999by that show that the 
supernova is intrinsically polarized.  SN 1999by has an observed, 
overall level of polarization of $\approx$ 0.3 to 0.8\%,
a rise of the polarization  $P$ redward of 6500 \AA, and a change
in polarization across the \SiII\ 6150 \AA\ feature of about 0.4\%.
The presentation of the polarization at different wavelengths
in the \qu\ plane is shown to be a powerful tool to determine 
the overall geometry and the interstellar component. 
The distribution of points with wavelength using this empirical 
\qu\ plane method reveals that SN~1999by has a well-defined 
axis of symmetry and suggests an interstellar polarization (ISP) vector 
with $P_{\rm ISP}=0.3\%$ and position angle $\Theta = 150\deg$ with 
an error circle in the \qu\ plane of radius about 0.1\%.

Synthetic NLTE-spectra for axisymmetric configurations based on 
delayed detonation models have been computed assuming ellipsoidal geometry.  
The input ejecta structure and composition are based on a 
Chandrasekhar-mass delayed-detonation model. 
The parameters of the explosion are chosen
to reproduce the time evolution of IR-spectra of SN~1999by 
without further adjustments.  Spherical models are then mapped onto
ellipsoidal geometries and the axis ratio, viewing angle and
interstellar polarization
adjusted to provide the best agreement with the polarization spectra.
Both flux and polarization spectra can be reasonably well reproduced by
models with an asphericity of $\approx 20 \% $ observed equator-on. 
The general properties of the polarization can be understood as a 
consequence of the structure of subluminous models. Best
fits are obtained for the theoretical models with $P_{\rm ISP}=0.25\%$ and  
$\Theta = 140\deg$, consistent with the empirical method.

We discuss our results for this subluminous Type Ia in the context 
of ``normally bright" Type Ia supernovae.  For normally-bright
Type Ia, the photosphere is near the inner iron-rich
layers at maximum light and the ubiquitous iron lines give a 
rapid variation to the model polarization spectra.  In subluminous 
models, the photosphere
near maximum is in the silicon layers with fewer lines and a
smoother overall polarization spectrum, as observed for SN~1999by.
Though data are sparse, the low upper limits for polarization determined 
for many normal events in contrast to the high polarization in SN~1999by 
may suggest a relation between the asymmetry we observed and the mechanism 
that produces a subluminous Type Ia.  Among various mechanisms, 
rapid rotation of the progenitor white dwarf, or an explosion during a 
binary white dwarf merger process are likely candidates to explain the 
asphericity in SN~1999by.

\end{abstract}


\section{Introduction}
The last decade has witnessed the explosive growth of high-quality 
data on Type Ia supernovae (SNe Ia) both from the ground and from space 
with spectacular results and new perspectives. One of the main advances was 
the confirmation of the long suspected correlation between the peak 
brightness and the decline rate (Phillips, 1993) which
clearly established that SNe~Ia span a wide range 
in brightness but, nevertheless, most
are found in a narrow range of brightness (Branch \& Tammann 1992). This
established SNe~Ia as a tool for accurate distance determinations on cosmological scales.
SNe~Ia have provided new estimates for the value of the Hubble constant (H$_0$)
based on a purely empirical procedure (Hamuy et al. 1996ab; Riess et al. 1996) and on a
comparison of detailed theoretical models with observations (H\"oflich \& Khokhlov 1996, hereafter HK96;
Nugent et al. 1997).
The values obtained ($H_0\approx65$ \Hunits ) are in good agreement with one another and
with those from the HST Key Project on the Hubble Constant based on Cepheid
distances. More recently,  routinely successful  detections of supernovae at
large redshifts (Perlmutter et al. 1997, 1999; Schmidt et al. 1998; Riess  et al. 1999)
have provided a new tool to measure cosmological
parameters such as $\Omega_M$ and $\Omega_\Lambda$. This has led to refined
estimates of H$_0$, indicated that we live in a low matter-density Universe, and,
most intriguing of all, gives hints that the cosmological constant is  not zero.

 Although details of the explosion are still under discussion,
there is general agreement that SNe~Ia result from some  process of combustion of a
degenerate white dwarf (Hoyle \& Fowler 1960). Within this general picture, three
classes of models have been considered: (1) explosion of a CO white dwarf (WD), with mass
close to the Chandrasekhar mass, which accretes mass through Roche-lobe overflow
from an evolved companion  star (Nomoto \& Sugimoto 1977); (2)
explosion of a rotating configuration formed from the merging of  two low-mass WDs,
caused by the loss of angular momentum due to gravitational radiation
(Webbink 1984; Iben \& Tutukov 1984; Paczy\'nski 1985); and
(3) explosion of a low mass CO-WD
triggered by the detonation of an outer helium  layer (Nomoto 1980; 
Woosley \& Weaver 1986;  HK96).
Only the first two scenarios seem to be in agreement with current observations.
Delayed-detonation models (Khokhlov 1991; Yamaoka et al. 1992; Woosley \& Weaver 1994) of
Chandrasekhar mass CO-WDs and their variations
can account for the spectral and  light curve evolution  of 
``normally bright" and subluminous SNe~Ia in the optical and IR 
(H\"oflich 1995a; H\"oflich et al. 1995a, hereafter HKW95; Wheeler et al. 
1998; Lenz et al. 1999;  H\"oflich et al. 2000). 
In this model, a burning front starts as a subsonic
deflagration and then undergoes a transition to a supersonic detonation.
Pure deflagration models like W7 (Nomoto et al. 1984) 
can also account for optical spectra of normal supernovae
(e.g. Nugent et al. 1997).  
The merging scenario remains an interesting alternative which to a great 
extent has
only been parameterized in theoretical studies 
but may produce results in agreement with some supernovae (HK96).
The sub-Chandrasekhar model triggered by edge-lit helium ignition
is disfavored on the basis of predicted light
curves and spectra (H\"oflich et al. 1996; Nugent et al. 1997).

 Despite the success of analyses of light curves and  spectra, 
these traditional methods do not provide any direct information on the
geometrical structure of the envelope.  Polarization does
provide information about the geometry in the scattering-dominated 
atmospheres of supernovae (van den Hulst 1957; Chandrasekhar 1960;
Shapiro \& Sutherland  1982;  H\"oflich 1991, hereafter H91; Jeffery 1991, 
hereafter J91).
This is because electron scattering causes the electric field vectors of
scattered photons to take on a preferential orientation.  For an unresolved
spherical scattering atmosphere, the electric vectors cancel, yielding zero
net polarization; however, in an aspherical SN, there will be increased
scattering from the wings which produces electric field vectors that are
not completely cancelled by scattering from the rest of the disk of the SN.
In this case the observer would measure a net polarization.

Polarization can thus provide unique information on the geometry of 
SNe~Ia that will give essential hints to understand the physics of 
thermonuclear explosions and the progenitor systems.  
Though historically the vast majority of computational models have
assumed spherical symmetry, for some of the scenarios we may expect strong
deviations from spherical geometry.
For instance, the deflagration front
has been found to be highly aspherical due to large rising plumes
(Khokhlov 1995, 2000; Niemeyer \& Hillebrandt 1995).  For delayed 
detonations, the transition
of the deflagration front into a detonation may occur
at a point rather than in a shell (Livne 1999). The coalescence of two white
dwarfs or an otherwise rapidly rotating progenitor could also produce
an overall, global asphericity.
Deviations from sphericity may  result in a direction
dependence of the emitted flux, and consequently may affect the use 
of SNe~Ia as distance indicators.

In this work, we report polarization measurement and interpretation
of the subluminous SN 1999by.
The observations are presented in \S 2 and the results in \S 3.  In \S 4 we
describe an empirical method to
analyse the data and to derive the interstellar polarization.
In \S 5, theoretical models for delayed detonations  are used to
analyze the observations.
We close in \S 6 with a discussion of the results for SN 1999by,
put our findings into context with other SNe~Ia and different explosion scenarios, and discuss the implications for the use of SNe~Ia as standard candles.

\section{Observations and Data Reduction}

\subsection{The Imaging Grism Polarimeter}
All observations were taken with the dual beam Imaging Grism Polarimeter (IGP)
at the Cassegrain focus of the McDonald Observatory 2.1m telescope.  
IGP is the Imaging Grism Instrument (IGI) combined with spectropolarimetry 
optics (Goodrich 1991).  IGI is a simple focal reducer with a grism that 
can be moved into the collimated beam.  Off-the-shelf optics are used 
which have been anti-reflection coated for $\lambda > 4000$ \AA\ . 
Polarization capability is provided by a modified Glan-Taylor
polarizing beam splitter and a rotatable waveplate.  The Glan-Taylor
prism is made up of two calcite blocks.  The ordinary ray (o-ray) 
is totally internally reflected, while the extraordinary ray (e-ray)
passes through, due to different indices of refraction.  
The waveplate acts as a halfwave retarder and rotates the plane of 
polarization of incoming light.  The halfwave plate is ``superachromatic'' 
with a retardance of 180\deg\ $\pm$ 2\deg\ from 3200 \AA\ to 11000 \AA.
The intrinsic polarization of IGP has been previously determined to be low,
$P < 0.1$, see Wang, Wheeler, \& \hof\ (1997a; hereafter WWH97) and 
Wang et al. (2000a).  This was confirmed with the
observation of null polarization standards.  More extensive information
about the data reduction techniques, standard stars, and observational setup 
is available in Howell (2000a). 

\subsection{Setup of the Instrument}
A 2.1\arcsec\ slit was used with the
6000 \AA\ grism, giving a resolution of about 14 \AA.  For all observations, 
the 85mm lens was used in conjunction with the TK4 $1024^2$ CCD, yielding 
a plate scale of 0.5\arcsec\ ($\sim$ 3.9 \AA) per pixel.
The spectra were wavelength calibrated with an argon lamp. 
Polarization standards were observed at least once per night.
The observing log is shown in Table \ref{log99by}.

\subsection{Data Reduction}
The data were reduced in IRAF, using tasks written by the authors.  
The reduction methods followed were those presented in Miller \etal\ 
(1988), Trammell (1994), and Howell (2000a).
For SN 1999by, two sets of data were taken each night with 
each ``set'' consisting of four 20--30 minute integrations at 
waveplate position angles of 0, 45, 22.5, and 67.5 degrees.
Polarization and flux standards were observed each night.
Since spectropolarimetry measurements require a high signal-to-noise, the
SN was observed over the course of three nights, and the data were 
combined.  The data presented here have been corrected for redshift,
though this correction is negligible ($z=0.00213$).

\subsection{SN 1999by}
SN 1999by was discovered independently by R. Arbour, South 
Wonston, Hampshire, England (Arbour \etal\ 1999), and by the Lick 
Observatory Supernova Search (LOSS; cf. Treffers \etal\ 1997, Li \etal\ 1999).
The SN was found on the outskirts (100\arcsec\ west and 91\arcsec\ north of the nucleus)
 of the well-known Sb galaxy NGC 2841.  According to the NASA/IPAC 
Extragalactic Database 
(NED)\footnote{http://nedwww.ipac.caltech.edu/index.html} the host galaxy
is a LINER and has a heliocentric radial velocity of $638 \pm 3$ \kps.
According to the Lyon/Meudon Extragalactic Database
(LEDA)\footnote{http://www-obs.univ-lyon1.fr/leda/home\_leda.html}, 
the velocity corrected for Virgo infall is 811 \kps.
Using $H_0=65$ \Hunits, this puts the distance to NGC 2841 at
12.49 Mpc, with a distance modulus of 30.48 mag.
This prodigious galaxy has also produced SN 1912A (type uncertain) and
the ``peculiar Type I's'' SN 1957A and SN 1972R.  

According to Bonanos \etal\ (1999), SN 1999by reached a maximum light of 
$B=13.80 \pm 0.02$ on UT 1999 May 10.5, and a maximum in the $V$ band of 
$V=13.36 \pm 0.02$ (date not given).  According to Weidong Li 
(private communication) the maxima were May 10.3 ($B$) and 
May 12.3 ($V$).  
Using a distance modulus of 30.48, and the photometry of Bonanos \etal , 
the absolute magnitudes at peak are $M_B=-16.68$ and $M_V=-17.12$.  
Thus SN 1999by
is underluminous by roughly 2.5 magnitudes compared to a typical
SN~Ia.  SN~1999by also had a large $\drp$ of 1.87.  This makes
SN~1999by nearly as steeply declining as SN~1991bg, which had $\drp = 1.93$.
SN~1999by is thus one of the most underluminous and rapidly declining
 SNe~Ia known.  SN~1999by is
similar to SN~1991bg in other respects as well.  The \SiII\ line at 
5800 \AA\ is strong relative to the line at 6150 \AA, another signature 
of the subluminous subclass.  According to Bonanos et al.
(1999) the flux near 4000 \AA\ was depressed as in other 
subluminous SNe~Ia, a feature thought to be due to \TiII\ absorption.  
The \OI\ feature at 7500 \AA\ is also strong in both SN~1991bg and SN~1999by.

\section{Results}
\subsection{Flux Spectra}
\label{fluxtext}
\begin{figure}[!htp]
\begin{center}
\plotone{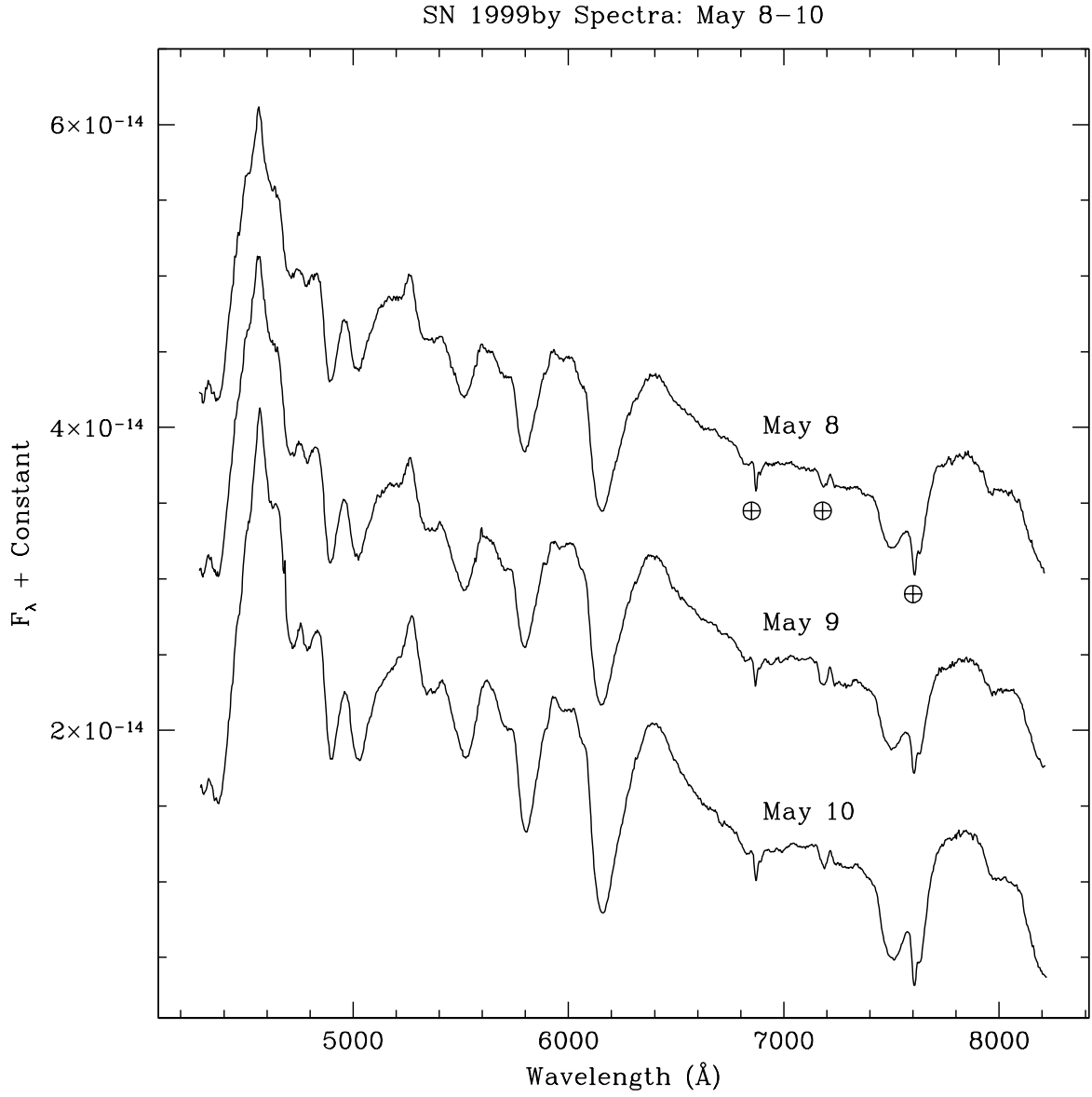}
\end{center}
\caption[Flux spectra of SN 1999by at $-2$ days, $-1$ day, and $B$ maximum.]{Flux 
spectra of SN 1999by at -2 days, -1 day, and $B$ maximum.  The data are in
arbitrary units.  Earth symbols denote terrestrial atmospheric features.}
\label{flux}
\end{figure}

Flux spectra of SN 1999by can be seen in Figure \ref{flux}.  Maximum light 
in $B$ occurred on May 10, therefore the spectra shown are two days before,
one day before, and at maximum light.  
Several features are apparent upon inspection Figure \ref{flux}.  Most
obvious is the fact that SN~1999by is unquestionably a subluminous SN~Ia.
It is similar to the prototype, SN~1991bg.           

Beginning at the blue side of the spectrum, we see that the spectrum
starts near 4300 \AA\ in a huge absorption feature.  This is thought to
be due to \TiII , a feature only observed in subluminous events 
(Filippenko et al. 1992, hereafter F92).  The 
fact that this ionization state of Ti is seen indicates a relatively low 
temperature in the ejecta, since \TiII\ is ionized in typical SNe~Ia 
(Nugent et al. 1995, hereafter N95).
\TiII\ is also responsible for absorption features in the 4500--5000 \AA\ 
region (F92). 
Moving redward, another signature of subluminous events is seen in the
pair of \SII\ lines on either side of 5400 \AA.  As one moves down the
sequence from overluminous to subluminous SNe, the blue line is overtaken
in strength by the red one (N95).  
Here the bluer line is scarcely seen at all,
consistent with our finding that SN 1999by is a very subluminous SN Ia.
 From the blueshift of the absorption features of Si, one can measure an
approximate velocity of this element (and thus its placement in the 
ejecta, since $v\propto r$).  For SN 1999by we measure $v(\rm Si)\sim 10,000$ \kps ,
comparable to speeds of $10,000$ and $10,600$ \kps\ reported for SN 1991bg by F92
and Leibundgut \etal\ (1993), respectively.  This is slightly lower than the average Si velocity
in a normally bright SN~Ia, $11,000$ to $13,000$ \kps, though velocities derived
from absorption minima of broad doublets as seen in the \SiII\ 6150 \AA\ 
feature are only accurate to $\sim$ 1000 \kps .

On the far red end of the spectrum, the absorption at 
7500 \AA\ is yet another hallmark of subluminosity.   
This line is due largely to a blend of \OI\ $\lambda$7774 and \MgII\
on the red wing, separated here by narrow atmospheric absorption.
This feature is not seen in the luminous SN 1991T, but increases 
in strength as one goes down the luminosity sequence to subluminous SNe~Ia (N95).  This \OI\ feature is also absent in SN 1998de, a fast-declining SN Ia 
that was peculiar in several ways (Modjaz \etal\ 2000).

\subsection{Polarization}

The dominant sources 
of opacity in an SN Ia, line scattering and electron scattering, can both 
change the polarization of incident flux.  Electron scattering 
polarization is independent of wavelength.  
In the case of axisymmetric scattering, the polarization position angle
can take on only one of two values: 0\deg\ or 90\deg\ with respect to the
symmetry axis (see \eg\ Cropper \etal\ 1988; J91, H91).
Line scattering need  not lead to complete angular 
redistribution.  The degree to which line scattering can polarize 
depends on the lower- and upper-level total angular momenta of that line (J91).
Despite the fact that lines can polarize, most line scattering is 
less polarizing than electron scattering by an order of magnitude.
As a result, electrons are the chief polarization mechanism, while the 
lines generally depolarize the flux previously polarized  by electron 
scattering.  Lines can redirect photons by decay in the same transition,
but they can also redistribute photons to both higher and lower frequencies
by further excitation or ionization from the level at which the photon is 
absorbed or by decay into multiple discrete lower levels 
(J91, H\"oflich 1995a).

Dust can polarize by scattering through dichroic extinction and through
emission of polarized photons.
Dust scattering is usually wavelength dependent, though the results
vary with scattering angle.  For dust typical of the ISM, the polarized flux
of optical photons undergoing a single scattering can decrease with
wavelength for small scattering angles, increase with wavelength
for large scattering angles, or remain constant for $\Theta_{sc}\simeq
90\deg$ (White 1979; Webb et al. 1993). 
Nonspherical dust grains with aligned rotation axes (from, e.g. a magnetic
field) can be a source of polarization through dichroic extinction.  In
this case, the polarization position vector is perpendicular to the grain
alignment vector.  

\begin{figure}[!htp]
\begin{center}
\plotone{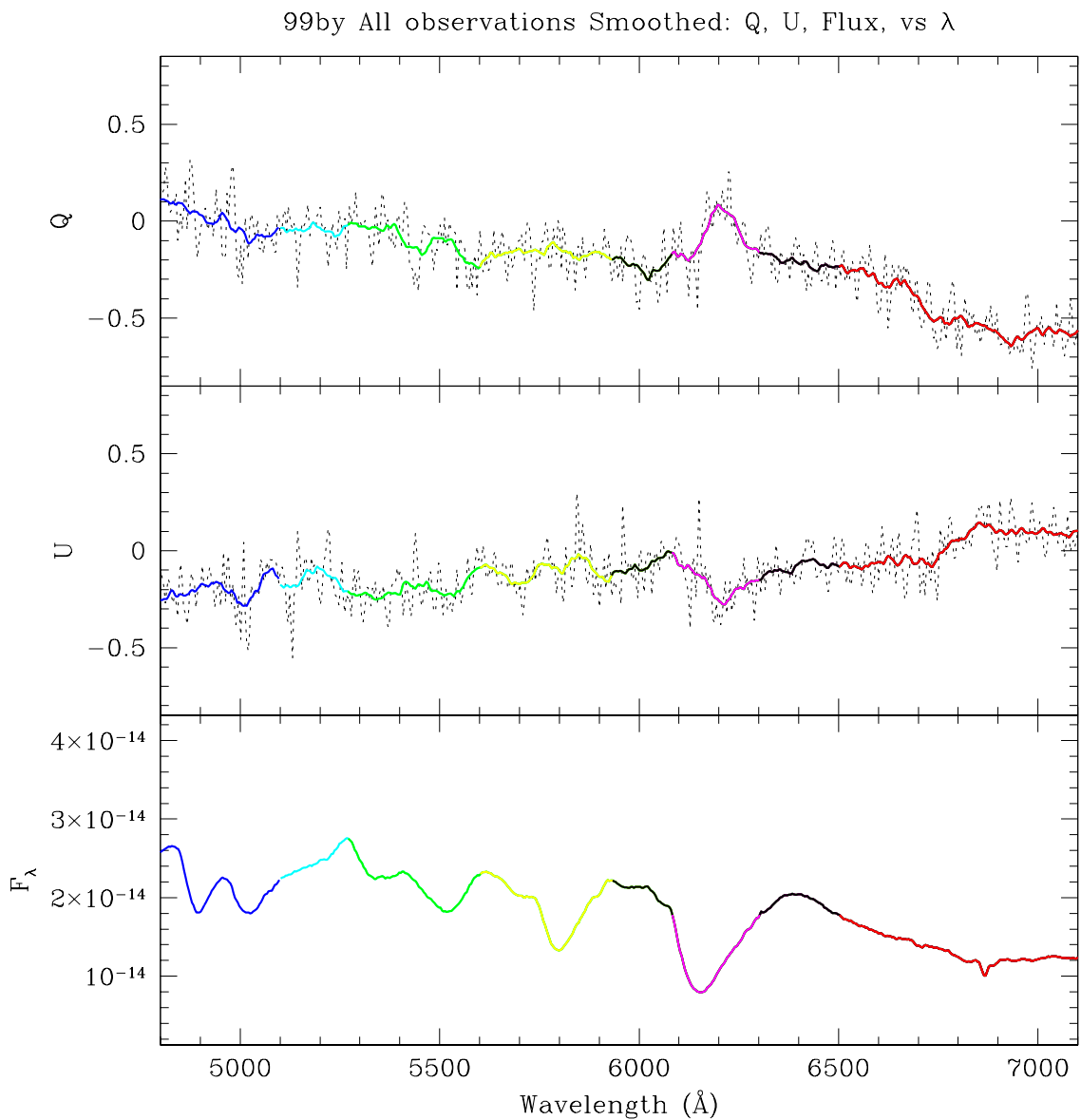}
\end{center}
\caption{Unsmoothed (dashed line) and smoothed, color coded $Q$, $U$, and 
flux spectra of SN 1999by created by 
summing all observations.  
Features from 4800--5100 \AA\ are blue, from 5100--5270 \AA\ light blue.  The \SII\
features from 5270 to 5600 are green.  The \SiII\ 5800 \AA\ feature is 
yellow, while the \SiII\ 6150 \AA\ line is magenta.  Black represents the 
area between the Si features and the area slightly redward.  The large, 
broad polarization feature to the red of 6500 \AA\ is colored red.
Colors are the same as in Figure~\ref{qutotrc}.}
\label{quftotrc}
\end{figure}

Table 2 gives all the SNe~Ia for which some polarization data are available.
Figure \ref{quftotrc} shows the normalized Stokes parameters $Q$ and $U$ 
and the flux spectrum of SN~1999by.  The dashed line is the unsmoothed data.  
A darker colored line on the $Q$ and $U$ plots shows the data smoothed 
with a running boxcar smoothing function of 17 pixels (65 \AA ).  
Features from 4800--5100 \AA\ are blue, from 5100--5270 \AA\ light blue.  
The \SII\ features from 5270 to 5600 are green.  The \SiII\ features are
marked by two different colors.  The 5800 \AA\ feature is yellow, while the
6150 \AA\ line is magenta.  Black represents the area between the Si features
and the area slightly redward.  Finally, the large, broad polarization
feature to the red of 6500 \AA\ is colored red.

\begin{figure}[!htp]
\begin{center}
\plotone{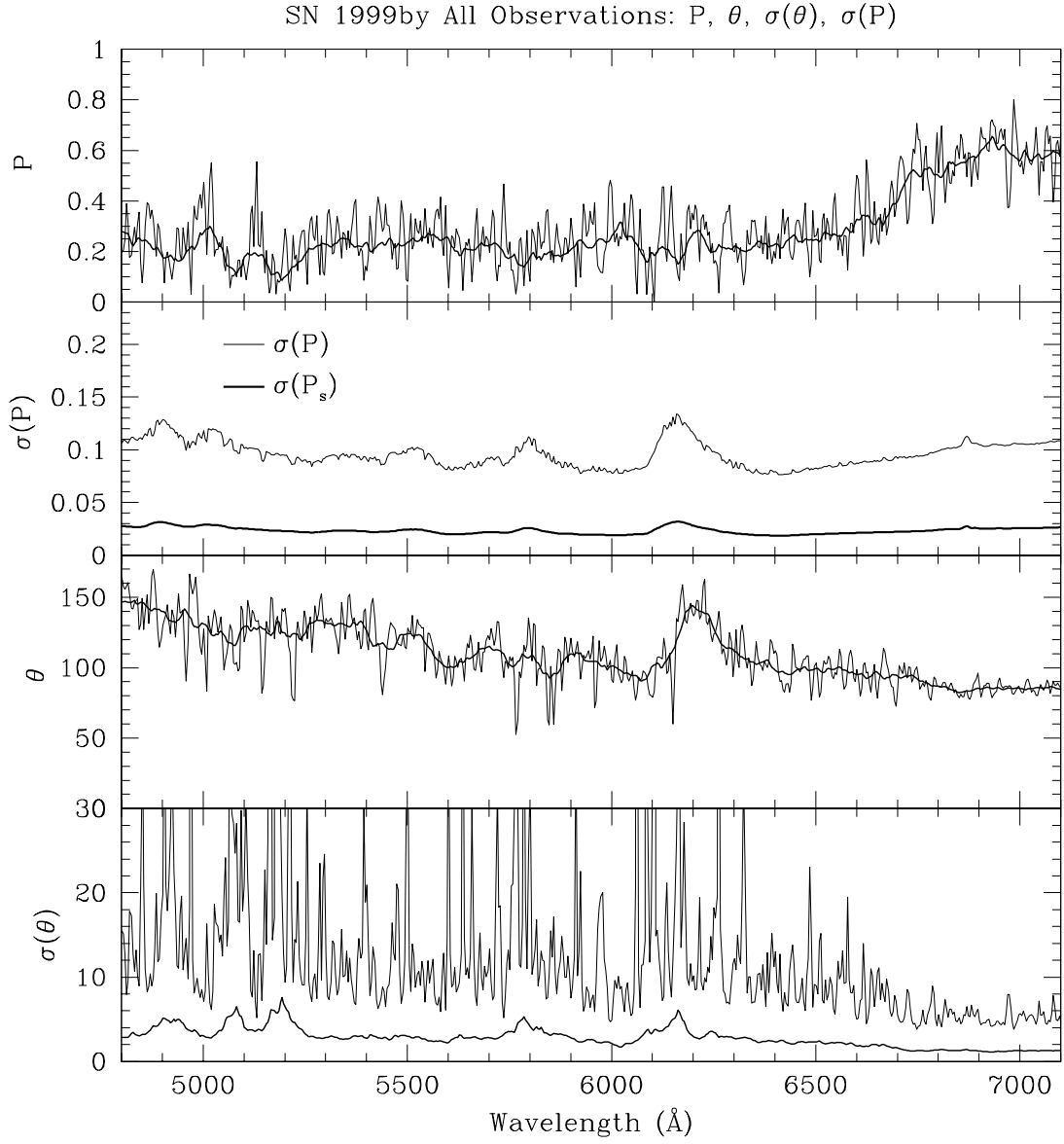}
\end{center}
\caption{$P$, $\Theta, \sigma(\Theta), \sigma(P)$ for SN 1999by created by
summing all observations.}
\label{tsstot}
\end{figure}

In Figure \ref{tsstot} we present $P=\sqrt{Q^2+U^2}$ for the
unsmoothed data, but we note that this indicator is biased high for low 
signal to noise data. 
Therefore, we generate the smoothed version, $P_s$ (dark solid line)
from the smoothed $Q$ and $U$ values.  In other words, 
$$P_s = \sqrt{Q_s^2 + U_s^2},\eqno{[1]} $$
where the subscript $s$ denotes smoothed values.  
Similarly, the smoothed polarization position angle is: 
$$\Theta_s=0.5 \arctan{(U_s/Q_s)}.\eqno{[2]}$$
In addition to $P$, Figure \ref{tsstot} also shows $\Theta$ and the 
associated errors $\sigma(P)$, and $\sigma(\Theta)$.

Figure \ref{qutotrc} shows the smoothed values for $Q$ and $U$
in the \qu\ plane for SN~1999by.  The colors are identical to the
color scheme used in Figure \ref{quftotrc}.  In such a plot, each point
defines a vector from the origin.  The length of this vector is $P_s$
and the orientation is $2\Theta_s$.  
Figure \ref{qutotrc} and its implications are discussed further in 
\S \ref{qudiscussion}.

As is often the case in spectropolarimetry measurements, the data for
any single set of observations has a low signal to noise.  Thus we have 
combined the data for all three nights to produce Figures \ref{quftotrc} and
\ref{tsstot}.  
These figure show the sum of all observations, and consequently
have the best signal to noise, with $\sigma (P) \sim 0.1\%$, and 
$\sigma (P_s) \sim 0.025\%$.  Note that despite
this low formal statistical error, systematic errors prevent us from actually
discriminating small scale  changes in $P_s$ of order 0.05 to  0.1\%.
We use this ``lowest noise'' 
case for most subsequent analysis, but we also look to the individual
data sets for confirmation that features are real.  Figure \ref{4p}
shows the results of combining the data in different ways.  
The top panel shows $P$ derived from all observations.  The second panel
shows $P$ constructed from only the best four data sets.
The third panel shows $P$ from the best single data set, taken on May 10
when the supernova was at maximum light.  Finally, the lower panel of
Figure \ref{4p} shows the sum of the best data from May 8 and 9 combined 
with the second best data set from May 10.  These are the next
three best data sets after the best set from May 10.
The dashed lines in Figure \ref{4p} are 1 $\sigma$ errors.  Our
indicator $P_s$ falls beneath the midpoint of the unsmoothed data
because $P$ is biased high as an indicator, though $P_s$ is less biased.
The degree of the bias to $P$ increases with low signal to noise data.
We also tested Wang's indicator,  (WWH97) and the rotated Stokes parameter
(\eg\ Tran 1995), and both agree with $P_s$ to within 0.02\%
across all wavelengths. 
It is important to remember that there is no perfect indicator for the total
degree of polarization (Leonard \etal\ 2000b; Simmons \& Stewart 1985).  
Three features stand out in every data set.  These are: the overall
level of $P \sim 0.25\%$, the change in $\Theta$ across the 6150 \AA\
\SiII\ feature, and the rise in $P$ to 0.6\% at 6900 \AA . 
We discuss each in turn.

\noindent{\bf The Level of $P$:}
The first of these significant features
is that  throughout most of the spectrum $P \sim 0.25\%$, with
some excursions to lower values.  
Not every bump in the smoothed data in Figure \ref{4p} is significant
because smoothing can introduce artifacts into low signal to noise data;
however, the overall level of
$P_s \simeq 0.25 \%$ shown in Figure \ref{tsstot} is significant.
The statistical error for $P$ is
$\sigma(p)\sim 0.1\%$ in the region of interest, and the statistical error
for the smoothed spectrum is $\sigma(P_s) \sim 0.025\%$.  
We should also keep in mind that the fact that 
we have smoothed the data means that we cannot distinguish features 
narrower than 65 \AA .

\noindent{\bf \SiII\ 6150 \AA :}
An overall level of polarization is interesting, but could be caused by
interstellar or circumstellar sources. 
Perhaps the most striking feature of the spectropolarimetry of SN 1999by
is then the change in position angle, $\Theta$, of the polarization through 
the \SiII\ 6150 \AA\ feature as seen in Figure \ref{tsstot}.  The PA jumps 
from 100\deg\ to 150\deg\ over this region, peaking at 6200 \AA.  As shown
in the second panel of Figure \ref{tsstot}, $\sigma(\Theta) \sim 12$\deg\ 
without smoothing.  After smoothing, $\sigma(\Theta) \sim 3$\deg .  
{\it Even without smoothing, the change in $\Theta$ across Si is a 4$\sigma$ 
result.}  This is clear evidence that the SN is intrinsically polarized.

It is also remarkable that the polarization change in Si is roughly 
equal to the polarization change across the entire wavelength region.
At 4800~\AA\ $\Theta \simeq 150$\deg\ but, by 7100 \AA\, $\Theta$ has
dropped to $\simeq 90$\deg\ .  This is significant because we will later
show that there is a single choice of interstellar polarization that, when
subtracted from the data, reduces the position angle across all 
wavelengths to nearly a single value.  

\noindent{\bf The 6900 \AA\ Rise:}
The rise in polarization toward the red peaks at $P \sim 0.6\%$ at 6900 \AA.
This feature is wider (1000 \AA ) than any particular line, and it occurs
in a region normally blanketed by Fe-peak lines.  Such a rise in polarization
in this wavelength region was predicted theoretically by WWH97.  We will 
revisit this feature when we discuss the theoretical interpretation of 
the data in \S \ref{theory}.

\section{Data Analysis Using the \qu\ Plane}
\label{qudiscussion}
The \qu\ plane is a powerful tool for interpreting SN polarimetry data 
both with respect to testing the overall geometry and to determining the 
interstellar component of the polarization.

\subsection{Intrinsic Properties of SN 1999by}
\begin{figure}[!htp]
\begin{center}
\plotone{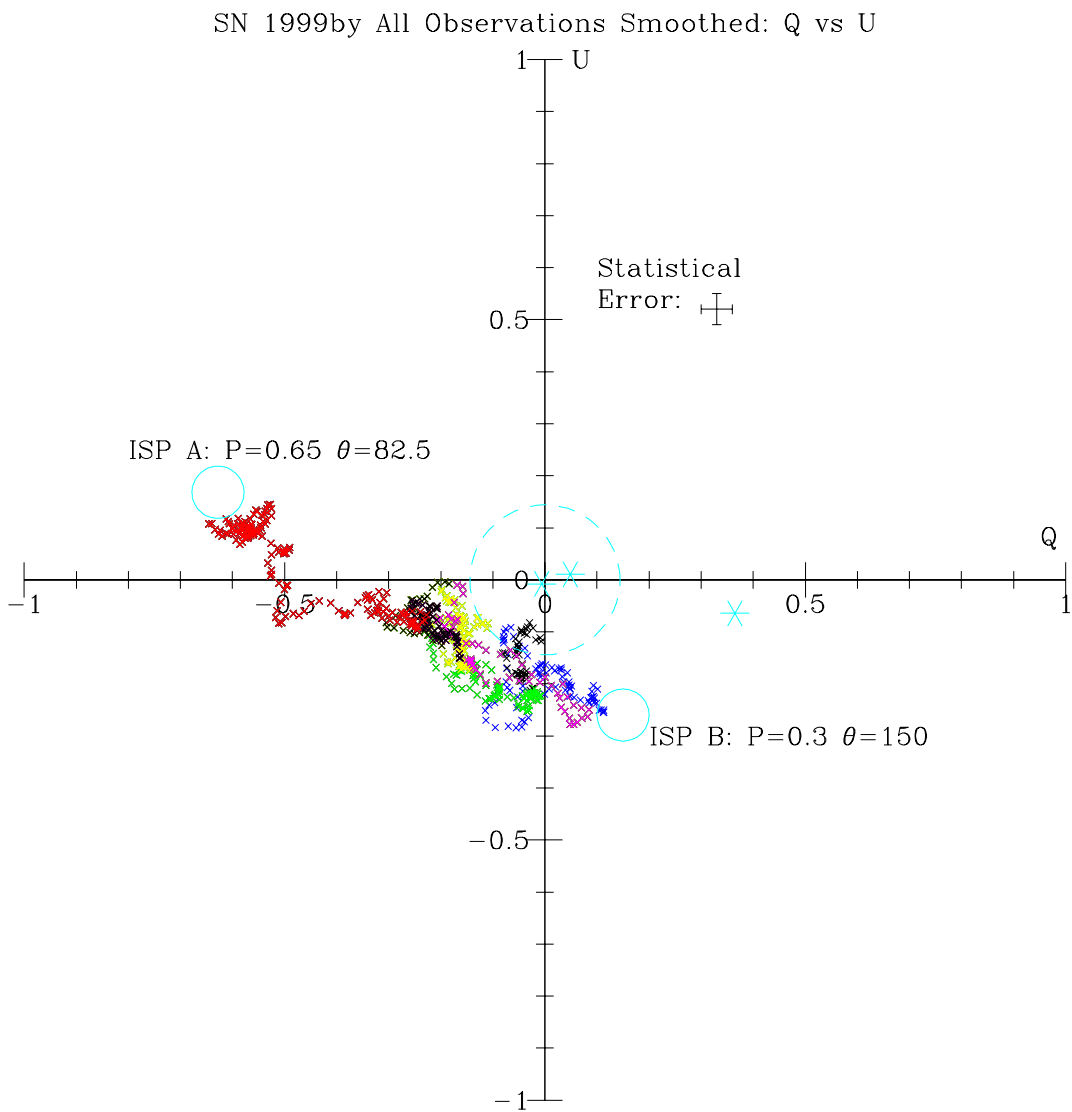}
\end{center}
\caption{Smoothed, color coded \qu\ plane for SN 1999by created by
summing all observations.  Each point corresponds to one pixel in the
spectrum.  Error bars are statistical error bars for
the smoothed data.  Light blue asterisks are polarization measurements of 
stars in the Galaxy within 5\deg\ of the line of sight of the SN.  
Light blue circles denote two choices of ISP 
discussed in the text.  The diameter of these circles corresponds to
the approximate uncertainty in the placement of the ISP using the empirical
method.  Dashed circle is the maximum allowable interstellar polarization 
from the Galaxy, as discussed in the text.  Colors are the same as 
in Figure~\ref{quftotrc}.
}
\label{qutotrc}
\end{figure}

\begin{figure}[!htp]
\begin{center}
\plotone{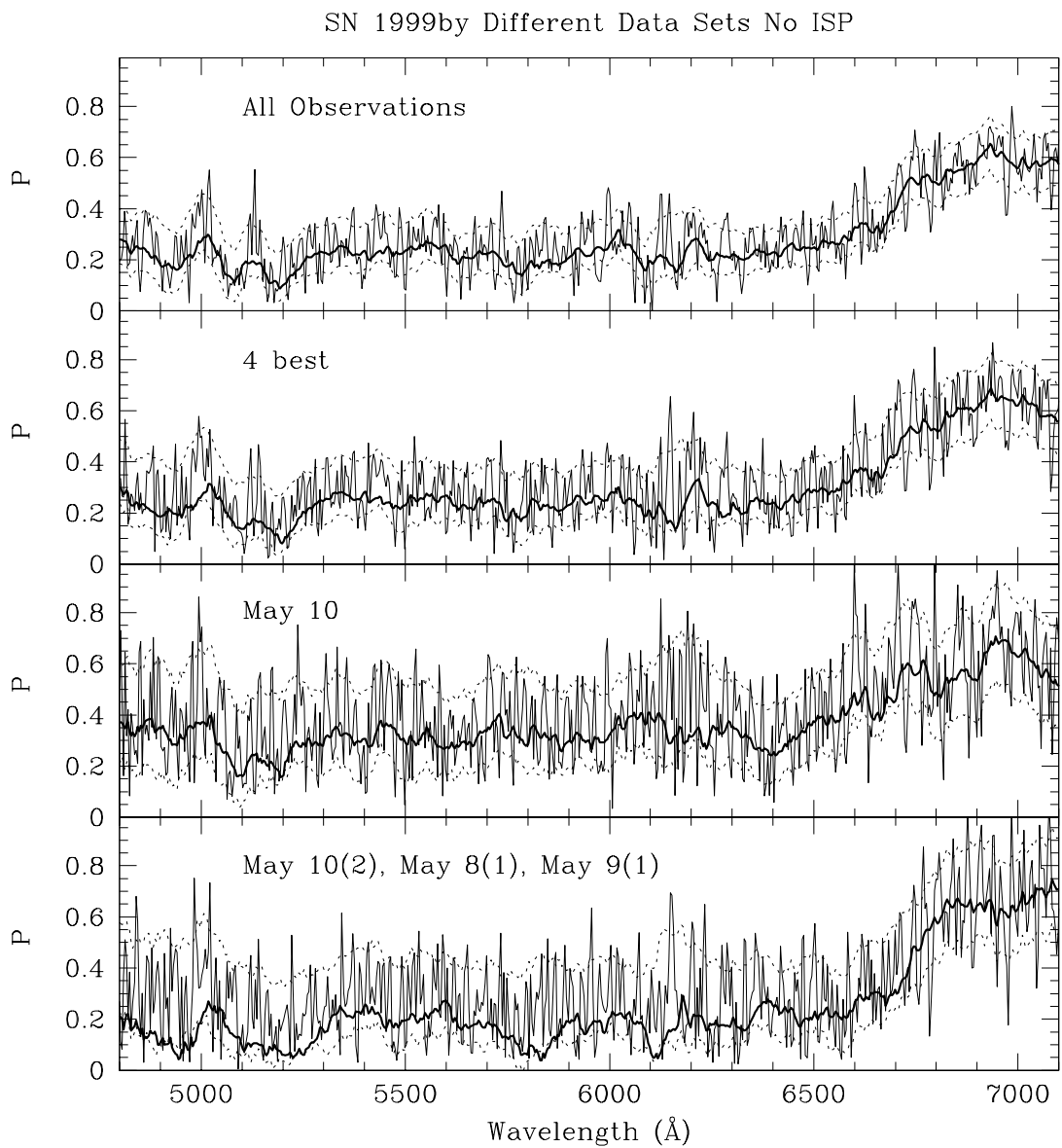}
\end{center}
\caption{Different data sets of $P$ for SN 1999by.  Top panel:  all observations
combined.  Second panel:  The 4 best sets of observations.
Third panel:  The best single data set --- the first data set taken on
the night of maximum light, May 10.  Last panel:  similar 
to the ``4 best'' case, but without the best spectrum (May 10-1).  This is 
the sum of the data sets May 8-1, May 9-1, and May 10-2 --- the 2nd through 
4th best sets.}
\label{4p}
\end{figure}

The \qu\ plot for the smoothed data of SN 1999by is given in Figure
 \ref{qutotrc}.  Each point corresponds to one pixel.  The points are 
color coded to reflect their distribution in wavelength.  The color coding 
is from Figure \ref{quftotrc} which shows the smoothed $Q$, $U$, and 
flux spectra.

Polarization arising from axisymmetric scattering will show up as 
a straight line on the \qu\ plot (see \eg\ Cropper \etal\ 1988; J91).
 Uncorrected for interstellar polarization, the
data for SN 1999by clearly define an axis of symmetry in 
Figure \ref{qutotrc}. The scatter is $\approx 0.1\%$, comparable to 
the scatter expected from photon statistics and systematic effects.

The distribution of the \SiII\ 6150 \AA\ feature is particularly
interesting, as noted in the previous section.   Here the Si feature 
is colored magenta.  It was previously noted that the variation in 
position angle across the Si feature matches the variation in position
angle seen across the entire wavelength region of study.   Stated another
way, \SiII\ is not localized to one spot on the \qu\ diagram, but
is distributed throughout the ``line'' demarked by the rest of the
points in the plot, except the points from the 6900 \AA\ region, which 
are colored red.  Physically, this means
that Si does not have a special geometry --- it shares the same 
axis of symmetry as the rest of the SN.

There is a slight deviation from linearity in the red points on
Figure \ref{qutotrc}.  If we take this jog to be real, 
noting that it is nearly perpendicular to
the rest of the points, then the jog corresponds to a switching from
an oblate to a prolate geometry in the corresponding wavelength 
region or vice versa.  This seems to be an unnecessary complication
of the model.  The discrepant points
can be explained by noise. If a point  deviates from
the line by 3$\sigma$, smoothing can introduce correlated errors,
causing several neighboring points to appear to deviate from the line as well.
On the basis of these few points, it is not possible to conclude that 
there are deviations from the global geometry in this wavelength region.  

\subsection{Interstellar Polarization}
\label{ISPtext}
Interstellar Polarization (ISP) due to dust in the interstellar medium
of both the host galaxy and the Milky Way can change the observed 
polarization spectrum of an object.  The difficult question of  
how much of the polarization signature measured for SN 1999by is 
intrinsic and how much is interstellar 
plagues polarization measurements of all objects.  The issue 
is particularly difficult in the case of extragalactic
objects which can have two sources of ISP --- one contribution from our
Galaxy, and one from the host galaxy.  Fortunately, we can 
estimate the ISP for SN~1999by via two essentially independent methods,
the empirical \qu\ plane method and theoretical considerations.  
We will discuss these specific
techniques below, but in this section we provide general constraints on 
interstellar polarization.

The maximum interstellar polarization allowable, corresponding to
the most favorable alignment of dust grains along the line of sight is 
$9\% \times E(B-V)$ (Serkowski, Mathewson \& Ford 1975).
The galactic longitude of SN 1999by is 166.91\deg\ and latitude 
44.12\deg\ (NED).  The Galactic extinction towards NGC 2841 is 
estimated to be $A_B=0.070$ mag (Schlegel \etal\ 1998) though 
Burstein \& Heiles (1982) give $A_B=0.000$ mag.  Schlegel \etal\ also 
estimate $E(B-V)=0.016$ mag.  Using Schlegel's 
estimate of $E(B-V)=0.016$ mag, the maximum Galactic component of 
polarization is 0.144\%.  This is indicated by a dashed circle in the
\qu\ plane diagram of Figure~\ref{qutotrc}.

Some limited information about the Galactic component of ISP can
be gleaned from observations of stars near the line of sight to SN 1999by.
There are only three stars with polarization measurements in the 
literature within $\pm 5$\deg\ of SN 1999by, and they are presented in 
Table \ref{polstars-table}.  They are also plotted as light blue asterisks 
on the \qu\ plane, Figure \ref{qutotrc}.
The closest star to SN 1999by is HD 82328.  It is 2.16\deg\ away from the 
SN on the sky.  While HD 82328 shows a low level of polarization, $P=0.01\%$,
it is also the closest star at 19 pc, so it does not sample much of the ISP 
between us and the supernova.  The next closest star, HD 82621 is 2.7\deg\ away
with $P=0.05\%$,
but it is only 40 pc away.  If we go out to HD 77770, we can look
through 1.3 kpc of the Galaxy to see $P=0.37\%$, but now we are looking
$3.25\deg$ away from the line of sight to the SN.  This is probably the
reason this star is outside the allowed Galactic ISP region of
Figure \ref{qutotrc}.  The three stars also have different polarization
position angles, and it is likely that any Galactic polarization toward
SN 1999by would have still a different position angle.  In addition,
if there is a contribution from ISP in the spectrum of SN 1999by, it could
arise from the host galaxy.  These three stars alone do not allow us to
place meaningful limits on the ISP.  Fortunately, the \qu\ plane does
suggest a choice of interstellar polarization.

\subsection{ISP Subtraction}

ISP is a smoothly varying function of wavelength, 
described by the Serkowski law:  
$$P/P_{\rm max}={\rm exp}[-K \rm{ln}^2(\lambda_{\rm max} / \lambda)]$$
where $p$ is the percent polarization at wavelength $\lambda$, 
$P_{\rm max}$ is the maximum polarization at wavelength $\lambda_{\rm max}$
and $K=0.01 + 1.66 \lambda_{\rm max},$ ($\lambda$ is in microns; Whittet \etal\ 1992).  

To remove ISP, we
subtract a Serkowski law from the data.  This is essentially a vector 
subtraction in the \qu\ plane.  Correcting 
for interstellar polarization can either increase or decrease the 
derived intrinsic polarization of the object, and it can turn absorption
features into emission features and vice versa.  This is easy to understand
if we remember that we are ``undiluting'' the light from the SN by 
removing contaminating ISP.  Mathematically, this is because the 
combination of the intrinsic polarization and ISP is a vector addition.

To remove the interstellar signature from our data, we must first 
determine $P_{\rm ISP}$.
As remarked above, the data points fall on a line in the \qu\ plane of
Figure \ref{qutotrc}. We can assume for SNe~Ia that
some parts of the spectra are depolarized (WWH97). Taken together, this
allows two possible
choices of interstellar polarization (see Wang et al. 2000b). 
If we place the ISP at
either end of the line delineated by the data points and subtract it,
then the data will fall along a radial line in \qu\ space with essentially one
constant value of $\Theta$ for the system (note that $\Theta$ is one-half
the angle of a vector on the \qu\ plane).  It is unlikely that the ISP
point could be placed well within the spread of unsubtracted SN 1999by points.
In this case, after subtraction the position angle would change abruptly
where there was no particular feature in the polarization spectrum.  We argue
that this behavior is not likely to be physical.

The two choices for ISP are the light blue circles marked as A and B in
Figure \ref{qutotrc}.  The diameter of the circles corresponds to the
approximate uncertainty in the placement of the ISP.  ISP A has a value 
of $P=0.65\%$ and $\Theta=82.5\deg$.  ISP B has a value of $P=0.3\%$
and $\Theta=150\deg$.
Both of the choices for interstellar polarization lie outside of that
allowed for the Galaxy along the line of sight to SN 1999by.
Assuming that the $E(B-V)$ of Schegel \etal\ (1998) is correct, this
implies that some of the ISP arose in the host galaxy.
Note that we could place the ISP any place on the line defined
by the data in the \qu\ plane that was not well within the spread of
data points.  The further the ISP from the data, the more tightly, but
arbitrarily, the ISP position angle would be defined.  We cannot rule out
more extreme choices, but argue that the ISP is most plausibly at either
A or B so that some portions of the data represent low polarization
in the supernova.  We think this argument applies to SNe~Ia, but note
that the data on the Type II SN~1999em show that at some phases there
is finite polarization at all observed wavelengths (Wang et al. 2000b).

\begin{figure}[!htp]
\begin{center}
\plotone{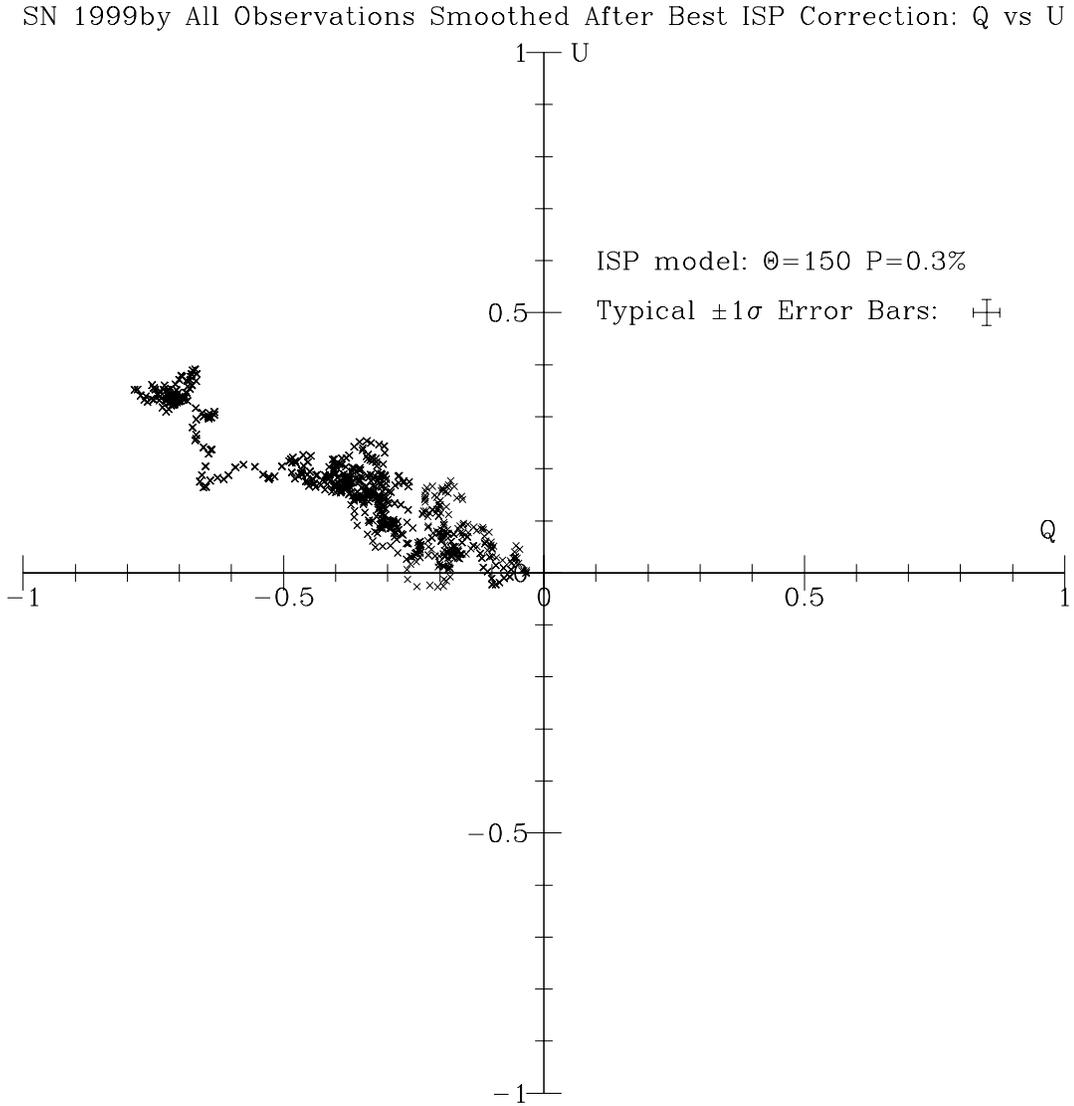}
\end{center}
\caption[All observations: $Q$, $U$, $P$ and $\Theta$ for SN 1999by after subtraction of ISP A.]{All observations: $Q$, $U$, $P$ and $\theta$ for SN 1999by after subtraction of $P_{\rm ISP}$ A.  The resulting polarization spectrum argues against this choice on theoretical grounds.}  
\label{quptaisp}
\end{figure}

\noindent{\bf ISP Choice A:}
First we consider ISP A.  Figure \ref{quptaisp} presents 
the resulting $Q$, $U$, $P$, and $\Theta$ with
this choice of $P_{\rm ISP}$ subtracted from the data. 
Note that in the lower panel, the effect of
removing the interstellar polarization is that the position angle 
becomes nearly constant across the entire wavelength region.  There is 
some deviation at longer wavelengths.  This is due to the fact
that when the ISP-corrected points are relocated on the \qu\ plane,
the long wavelength points are closest to the origin, and deviations 
in position angle are magnified.
ISP A is disfavored for three reasons:
it produces a decrease in polarization at 7000 \AA, it produces an
increase in polarization in the \SiII\ 6150 \AA\ feature, and it is
less likely due to reddening considerations.  We consider each in turn.

Subtracting ISP A produces a high
$P$ $\sim$ 0.8\%, at short wavelengths, and a decrease in polarization at
wavelengths greater than 6500 \AA.
This is contrary to theoretical expectations.  Numerical calculations
presented in WWH97 predicted a {\it rise} in polarization
longward of 6500 \AA\ due to decreased line blanketing.  This
is discussed in further detail below in \S\ref{theory}. 

The removal of ISP A from the data also causes the polarization to
be enhanced in the Si 6150 \AA\ feature.  While this increase cannot be ruled out
completely, from a theoretical perspective, this feature is easier to
understand if it is a depolarization.  P-Cygni lines are generated
above the photosphere, so polarization produced by electron scattering
at the photosphere is expected to be depolarized by absorption and reemission
in the layers above it (see the appendix).  Note that by using this 
qualitative theoretical prediction to discriminate between the two 
choices of ISP, the ``empirical method'' is not completely independent
of theory.  However, the approach is significantly different from the
determination of ISP from a quantitative comparison to a specific 
theoretical model as done in \S 5 below.

Another consideration that slightly disfavors the choice of ISP A 
is the observational constraint that the SN does not appear to be reddened.
As mentioned above, the maximum interstellar polarization allowable, 
corresponding to the most favorable alignment of dust grains along the 
line of sight is $9\% \times E(B-V)$. We can use
this to place a lower limit on the reddening expected from our choice
of ISP.  Inverting, we have: $$E(B-V) \geq P_{\rm ISP}/9\% \eqno{[4]}$$
Thus for model A, $E(B-V) \geq 0.072$, and for a standard extinction
law (Savage \& Mathis 1979), $A_V \geq 0.22$.  Given that SN 1999by 
is already bluer than SN 1991bg, which was thought to have low 
extinction along the line of sight, it is likely that SN 1999by is not
significantly reddened.  SN 1999by has a pseudocolor of 
$B_{\rm max} - V_{\max}$=0.44--0.47 (Bonnanos 1999; Li, private 
communication).  This compares to a value of 
$(B-V)_{\rm max}=0.74$ for SN 1991bg (M97).  Note that 
$B_{\rm max} - V_{\max}$ and $(B-V)_{\rm max}=0.74$ are not, strictly 
speaking, the same quantity, but this is the closest comparison we can make
until the photometry data is published.
Both SN 1991bg and SN 1999by are redder than typical SNe~Ia, but this is 
thought to be intrinsic to subluminous SNe~Ia (F92; Leibundgut et al. 1993; 
HKW95; Turatto \etal\ 1996; Mazzali \etal\ 1997).
Since neither the extinction towards SN 1991bg nor towards SN 1999by is 
known with certainty, and 
we can only place a lower limit on the reddening expected from our model, 
this is not a hard limit. 

\begin{figure}[!htp]
\begin{center}
\plotone{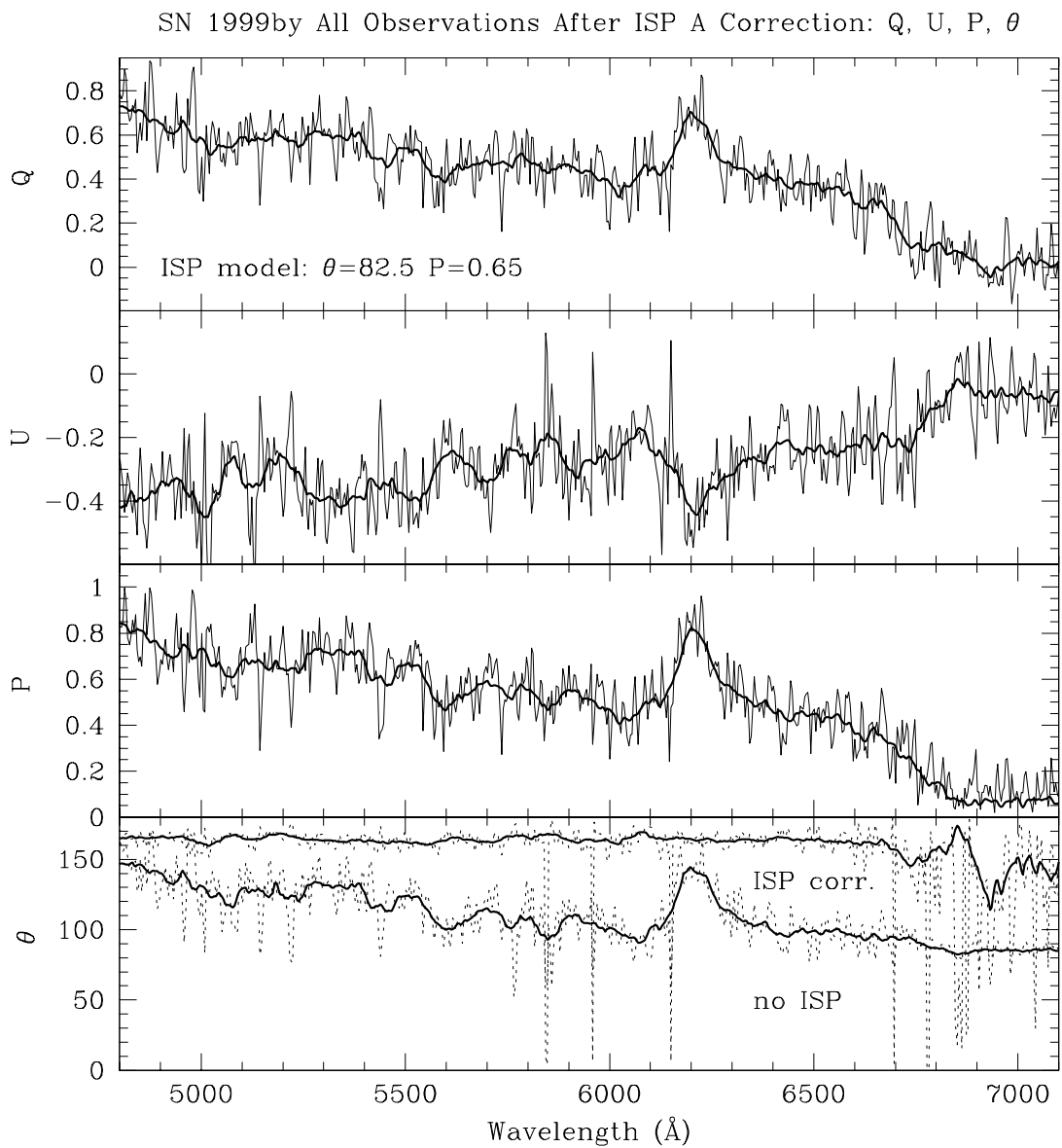}
\end{center}
\caption{The data in the \qu\ plane as in Figure~\ref{qutotrc} after 
correction for $P_{ISP B}$ with 
$P_{\rm ISP}=0.3\%$, $\Theta_{\rm ISP}=150\deg$.  This ISP has been 
vector subtracted from each point in accordance with the Serkowski law.  
The resulting distribution of points for SN 1999by is 
consistent with a constant polarization position angle, $\Theta=80\deg$ 
across all wavelengths.
}
\label{qutotrcisp}
\end{figure}

\begin{figure}[!htp]
\begin{center}
\plotone{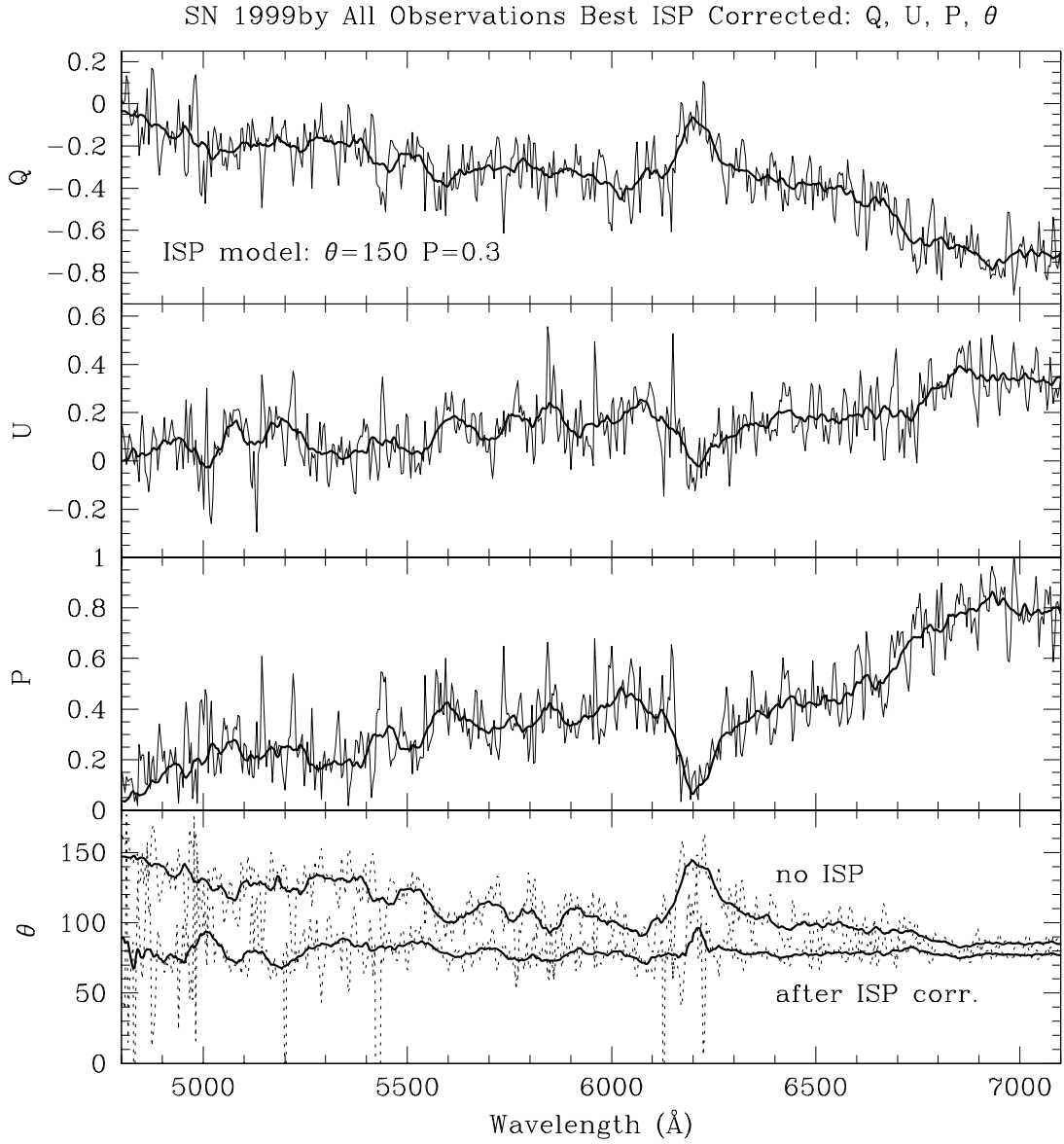}
\end{center}
\caption{All observations:   $Q$, $U$, $P$ and $\Theta$ for SN 1999by corrected for ISP B.  Note that after ISP correction, $\Theta$ is nearly constant 
with wavelength.}
\label{quptbestisp}
\end{figure}

\noindent{\bf ISP Choice B:}
The subtraction of ISP B from the data, like ISP A, causes the 
SN data to lie essentially on a straight radial
line in the \qu\ plane as can be seen in Figure \ref{qutotrcisp}.
For this choice of ISP, the polarization deduced for the SN has a 
constant position 
angle of $\Theta \simeq 80\deg$ across the entire wavelength region.  
In the bottom panel of Figure \ref{quptbestisp}, the 
ISP subtracted data is compared to the position angle
from the non-ISP-corrected data.  This figure also shows 
smoothed and unsmoothed versions of $P$, $Q$, and $U$ after ISP correction.
ISP B is slightly favored due to reddening considerations.  For this
model $E(B-V) \geq 0.033$, and $A_V \geq 0.10$, less than half that 
of ISP A.  From a theoretical perspective, this choice also gives 
the desired depolarization of the Si feature.  ISP B also gives a rise
in polarization at red wavelengths as theoretically predicted.
For these reasons it is a better choice than ISP A.

With ISP B subtracted, we find $P$ $\sim$ 0.4\% from 5600 -- 6600 \AA\, 
with the exception of the Si feature.  The difference between 
the polarization maximum at 6900 \AA\ and the overall level, 
(e.g. at 5800 \AA ), is 0.4\%.  This relative difference does not
depend on the choice of ISP given the assumption that 
intrinsic polarization defines a radial line on the \qu\ plot.
The maximum degree of polarization after subtraction of ISP B is
$P=0.8\%$.  As noted above, subtraction of ISP A gives the same
maximum value, but for ISP B
the maximum polarization occurs at wavelengths $\ge 6500$ \AA .  This
is where the maximum polarization is expected from theory (WWH97).
The subtraction of ISP B produces relative
depolarization around 4800--5600 \AA, where the level is 0.2\% or lower.  
This is where line blanketing opacity dominates electron 
scattering opacity and thus is where depolarization is 
expected from theory (H\"oflich, M\"uller \& Khokhlov 1993, hereafter HMK93; WWH97).

We have estimated the degree of uncertainty associated with this choice
of ISP by choosing different values of $P_{\rm ISP}$ close to ISP B and 
observing the degree to which they cause the axis ratio of the system
to deviate from a single value across all wavelengths.  These are shown in 
Howell (2000a), but are omitted here for brevity.
After consideration of the uncertainties, our best choice of ISP, Case B, 
is $P=0.3\pm 0.05\%$ at $\Theta=150\pm 5\deg$.  

Regardless of whether we choose ISP A or B, the intrinsic polarization
of SN~1999by, given by the total length of the distribution of points in the \qu\ plane,
is $P = 0.8\%$.  As mentioned earlier, Si is distributed throughout
most of the line in the \qu\ plane and hence samples this total intrinsic polarization.  
One implication is that the Si 6150 \AA\ feature will always show
a minimum or maximum in the polarization after subtraction of a reasonable
choice of ISP lying at one or the other extreme of the data in the \qu\ plane.

\section{Comparison with Theoretical Models}
\label{theory}

The chemical and density structures for the initial models
are based on calculations of the stellar evolution from the
main sequence to white dwarf formation and the subsequent
phase of accretion onto the white dwarf (Umeda et al. 1999; H\"oflich, et al. 2000).
Spherical dynamical explosions and corresponding light curves are
first calculated.  Aspherical structures are then constructed 
by mapping the spherical structures in the homologous expansion
phase onto ellipsoidal density structures with an axis ratio $a/b$.
Here $a$ is the equatorial major axis and $b$ is the polar major axis.
Subsequently, detailed non-LTE spectra are calculated for the flux 
and polarization. Details are given in the appendix.

 Delayed detonation (DD) models and pulsating delayed detonation models 
have been found to reproduce the optical and infrared light curves 
and spectra of SNe~Ia reasonably well (H\"oflich 1995b; HKW95; HK96;
Nugent et al. 1997; Wheeler \etal\ 1998; Lenz et al. 1999; \hof\ \etal\ 2000;
Gerardy \etal\ 2000).  In particular, the DD models
with a variation in the parameter, $\rho_{tr}$, at which the
transition is made from deflagration to detonation, give a
range of ejecta structures and mass of \Ni56.  Models with smaller
transition density give less nickel and hence both lower peak luminosity
and lower temperatures. The latter gives lower opacities and hence
a steeper decline in the light curve.  The DD models thus give a
natural and physically well-motivated origin of the brightness-decline rate
relation of SNe~Ia within the paradigm of thermonuclear combustion of
Chandrasekhar-mass CO white dwarfs (HKW95; H\"oflich et al. 1996).

Here we present the polarization spectra of a DD model 
with parameters which have been adjusted to provide a fit 
to the time evolution of the infrared spectra observed for SN~1999by 
(Gerardy et al. 2000) without further tuning.  
In the delayed detonation scenario, the transition
density, $\rho_{tr}$, is the dominant factor that determines
the amount of \Ni56\ produced and hence 
distinguishes the normally-bright from the subluminous models 
(HKW95; H\"oflich 1995a; Iwamoto et al. 1999). 
We contrast the ejecta structure of a normally-bright
SN~Ia model with that required to reproduce a subluminous event 
like SN 1999by in Figure \ref{model1}.  The same progenitor structure
and central density at the time of the explosion have been taken
as the normally-bright SNe~Ia studied in H\"oflich et al. (2000). 
We assume a WD with a Chandrasekhar mass and solar metallicity 
which originates from a main sequence star of 7 $M_\odot$ (Umeda et al. 1999). 
At the time of the explosion, the central density is 
$\rho_{c}=2 \times 10^9$ g~cm$^{-3}$.
To produce a subluminous supernova consistent with the IR-spectra of SN~1999by,
$\rho_{tr}$ has been chosen to be $ 1 \times 10^7$ g~cm$^{-3}$
compared to $\rho_{tr} = 2.5 \times 10^7$ g~cm$^{-3}$ for
the normally-bright model. The subluminous model produced $0.103 M_\odot$ of
\Ni56\ compared to $0.701 M_\odot$ for the normally-bright model. 
A typical feature of subluminous DD  models
is the greater production of O, S and Si 
($0.251, ~0.431$ and $0.221 M_\odot$, respectively) compared to 
normally-bright SNe~Ia ($0.065, 0.215 $ and $0.123 M_\odot$, respectively). 
Another feature of subluminous SNe~Ia models is 
that the \Ni56\ is constrained to the inner layers with 
low expansion velocities which become visible a few weeks after maximum
light. In contrast, for normally-bright SNe~Ia models, 
the outer edge of the \Ni56\ layers extends to
about 10,000 to 13,000 km~s$^{-1}$ so that the Ni is
already visible at maximum light (Figure \ref{model1}). 
As we will see below, this qualitative difference between 
normally-bright and subluminous SNe~Ia has
important implications for our understanding of the polarization spectra.
For more details on the explosion models, light curves and spectral evolution,
see H\"oflich et al. 2000 and Gerardy et al. (2000).

 For this paper, we calculated the polarization spectrum at day 15 after the 
explosion which corresponds to maximum light in B.  At this epoch the model 
gives $M_V=-17.52^m$, and $B-V=+0.46$.  The temperature structures for the 
aspherical models were based on the spherical calculations. 
The optical spectrum is formed  in layers with expansion velocities between
$\approx 8000 $ and $ 14,000 $ \kps\ which corresponds to about  
1 $M_\odot $ in mass coordinate, \ie\ in layers
dominated by intermediate mass elements (Figure \ref{model1}). 
We calculated the spectra for a variety of ellipsoidal
structures at the photosphere with axis ratios $a/b$ between 1.15 and 1.5 
and adjusted $a/b$ to fit the level of polarization observed in SN~1999by 
for various inclination angles. 

 Currently, three-dimensional models are not a suitable tool to 
adjust parameters such as the main sequence mass, accretion rates,
metallicity, rotation and combustion physics to find best fits. Therefore,
we restrict our discussion of the fluxes to the main spectral features.

In our model, the spectrum at day 15 is formed at the outer edge 
of the Si-rich layers, i.e. well above the region containing 
\Ni56\ (Figure \ref{model1}).  The top panel of  Figure \ref{model2} shows
reasonable agreement between theoretical and observed features, 
and thus confirms the validity of this model in terms of
Doppler shifts of the lines, their  strengths, the ionization stages, 
and the overall slope of the spectrum. 
Most of the strong features can be attributed to the 
intermediate mass elements \OI , \SII\ and \SiII . 
Most of the weaker lines are due to iron-group elements which are singly
ionized as expected for subluminous SNe~Ia with a relatively 
cool photosphere (H\"oflich et al. 1995b).
The \OI\ feature at about 7450 \AA\ is too weak in the models. 
This is not due to a lack of oxygen (Figure \ref{model1}),
but the strength of the \OI\ line depends sensitively on the excitation. The
discrepancy may be an artifact of the approximations imposed by the numerical 
treatment, or it may hint at the need for a slightly higher excitation in 
the outer layers, \eg\ due to non-thermal electrons or an increased 
$\gamma $-ray flux.  We note that some fast-declining (subluminous) SNe Ia
do not show strong \OI\ in their spectra (Modjaz \etal\ 2000).

 The flux spectrum depends on the inclination, $i$ 
(here taken to be the angle between the equatorial plane and the line 
of sight).  For oblate geometries, the flux
increases with $i$ because of the increased escape probability of photons 
towards polar directions
(H91). Although the asphericity is small in our
example, the absolute flux varies by about 30\% or $0.3^m$
from equator to pole.
 With increasing $i$, weak absorption features
(e.g. due to \FeII , \SII) tend to be smeared out, 
and the overall spectrum  becomes smoother as shown in
the top panel of Figure \ref{model2}.
These absorption features are formed in a narrow region close to 
the origin of the quasi-continuum
which is produced by electron scattering, bound-free processes,  
and by a large number of very weak lines (H\"oflich 1995a). 
Because $ v\propto r$, for an oblate geometry the photosphere seen by an 
observer spans a wider velocity range as $i$ increases.  In addition, 
the probability of multiple scattering of photons tangential to the photosphere
into the ``absorption" feature increases with $i$ 
in the case of oblate geometries, thus diluting the depolarization of the line.

Flux and polarization spectra of a model with 17\% asymmetry 
observed at various inclination angles
are shown in Figure \ref{model2} in comparison with  observations.
The model polarization spectra have been rebinned with 
$\Delta \lambda = 12.5 $\AA, to provide a statistical error comparable 
to the observations which, for this section, have been rebinned
with $\Delta \lambda = 45 $ \AA. The model polarization spectra
for various choices of the inclination, $i$, and the raw and binned 
data are given in the second panel of Figure \ref{model2}.  The statistical
errors in the model and observations and the polarization angle before 
and after correcting for the model-determined ISP are given in the third 
and fourth panels of Figure \ref{model2}, respectively.

In the models, the polarization is produced by electron scattering. 
As discussed in the appendix, we assume complete redistribution for lines 
which causes depolarization in lines.
As expected both from analytical  and numerical studies,
the polarization $P \propto \cos^2 i$ (e.g. van den Hulst 1957; H91).
The frequency dependence of the polarization is governed by the 
absorption probability in lines relative to  electron scattering. 
Most prominent is  the strong depolarization by the \SiII\ line at  6150 \AA\ 
and the  \OI , \MgII\ and Si II features at 7600 \AA\ 
(Figure \ref{model2}, panel 2).  At shorter wavelengths, 
a large number of overlapping lines due to iron-group elements is
responsible for the low polarization. 
The wavelength region above 5400 \AA\  shows an increasing
degree of polarization due to the decreasing importance of line opacities. 
Shortwards of the \SiII\ line at 6150 \AA, the polarization level 
in the models is generally lower and shows modulations
due to moderately strong \SiII , \SII , and Fe~II lines, 
as  expected from the opacity pattern
for Si-rich layers at temperatures between 5000 to 10000 K (see Figure 1 in HMK93).
 
We note that the frequency pattern of the model polarization
spectra for subluminous supernovae is rather different from
normally-bright SNe~Ia at the same phase (e.g. SN 1996X, WWH97).
 In the latter, the polarization spectra near maximum light
are dominated by iron-group elements because the photosphere is between 
 10,000 and 12,000 km~s$^{-1}$, \ie\ right at the interface between
complete and incomplete Si burning (Figure \ref{model1}). 
In addition, the photospheric temperatures are
 higher than in subluminous events by several thousand degrees and, thus,  iron-group elements are present both
in the second and third ionization stage. 
The presence of these ubiquitous iron lines gives a more pronounced rapid frequency
dependence of the model polarization spectra for the normally-bright
SNe~Ia near maximum as contrasted with the relatively smooth
variation predicted for the polarization spectra of the subluminous
events.  

We have iterated the component of the interstellar polarization to 
optimize the fits with respect to the line width and depth of the 
strong lines, in particular the \SiII\ feature at 6150 \AA.  
The best agreement is for $P_{\rm ISP} =0.25 \%$ and $\Theta = 140\deg$.
Small changes of the ISP-vector result in slightly broader
components of the depolarization features. 
With increasing deviations from the optimized values,
local maxima appear in the line centers which eventually dominate the spectra.
 The vector for the interstellar polarization derived in
this way is consistent with the contribution of the ISP found  
with the empirical method described in \S 4.3.
The agreement supports the soundness and relibility of 
both the empirical and theoretical approaches.
 
As shown in Figure \ref{model2} (panel 2),
the observed polarization spectrum can be reproduced by an ellipsoid 
with an axis ratio of 1.17 seen equator on ($i \approx 0\deg$). 
The overall level of polarization in different wavelength
ranges and the velocity shift and strength of features produced by  
strong lines are consistent.

The theoretical polarization spectrum shows a distinct 
physical pattern between 5400 and 6000 \AA\ at the 0.1\%
level which cannot be discriminated in the data. 
These small scale fluctuations may be
valuable tools to analyze small scale structures
caused by inhomogeous mixing if better data become available in the future.

An axis ratio of 1.17 is a lower limit to the
amount of asphericity that is required, though we can also estimate 
an upper limit.
As an alternative to a model with small asphericity seen equator on, 
models with larger asphericity may be able to reproduce the 
observations if seen from larger inclinations, $i$.
To test this possibility, we have calculated a set of 
models with larger axis ratios as presented in Figure \ref{model3}.
To reproduce the overall level of polarization for $i$ of $28\deg$
and $52\deg$, the axis ratio $a/b$ must be boosted to 1.25 and 1.5, 
respectively.  Strong discrepancies at the $0.3 \% $-level
are present for $i=52\deg$ which are well beyond the level of
uncertainty. Most noticable are the strong peaks in $P$ at about 
$6000 $ and $6400 $ \AA, and the change of $0.6 \%$ 
between $5800$ and $6000$ \AA.  Such variations cannot be eliminated by 
a different choice for ISP.
  For $i=28\deg$ and $a/b$=1.25, the agreement with the observation of 
$P$ is better, but problems remain on the level of $0.2\%$, 
and there are some problems with the location of the \SiII\ minima. 
The model with $i=28\deg$ may be marginally consistent with the observations.
We can conclude that SN 1999by has intrinsic asymmetries of the order 
of $\approx 20 \%  $ and that it is seen almost equator on.

\section{Discussion and Conclusions}

 Whereas core collapse supernovae are commonly found to be polarized at  
1\% or greater, the degree of polarization is much smaller in 
SNe~Ia (Wang et al. 1996, 2000a).
Spectropolarimetric and broadband measurements of polarization in
SNe Ia are still exceedingly rare.  For most measurements, the interstellar
component cannot be determined, making any intrinsic component impossible to
determine.  Table \ref{polSNe-table} shows all SNe~Ia for which polarization 
data are available.  McCall et al.\ (1984b) measured no
significant polarization intrinsic to SN 1983G, but did place (large)
upper limits on spectropolarimetric signatures.  
The first slightly subluminous SN~Ia with a polarization measurement 
was SN 1986G.  Broadband UBVRIJH polarimetry with a 
maximum polarization of 5.2\% is well fit by a Serkowski law 
(Hough \etal\ 1987).
Broadband polarimetry of SN 1998bu is presented in Hernandez \etal\ (2000).
The polarization, ranging from 1--2\% again follows the Serkowski law. 
Spyromilio \& Bailey (1993) placed limits on the
spectropolarimetry of  SN 1992A which was observed
about 15 days and 100 days past maximum light.  
The overall level of $P$ is roughly 0.3\%, comparable to the noise level.
While Spyromilio \& Bailey make no claim for intrinsic polarization of the SN, 
polarization signatures of the type suggested for SN 1996X (WWH97) cannot
be ruled out from the published data.
Leonard \etal\ (2000a) presented spectropolarimetry of SN 1997dt, showing
apparent changes in $P$ across spectral features.  The level of 
intrinsic versus interstellar polarization is not known.

\begin{figure}[th]
\includegraphics[width=9.5cm,angle=270]{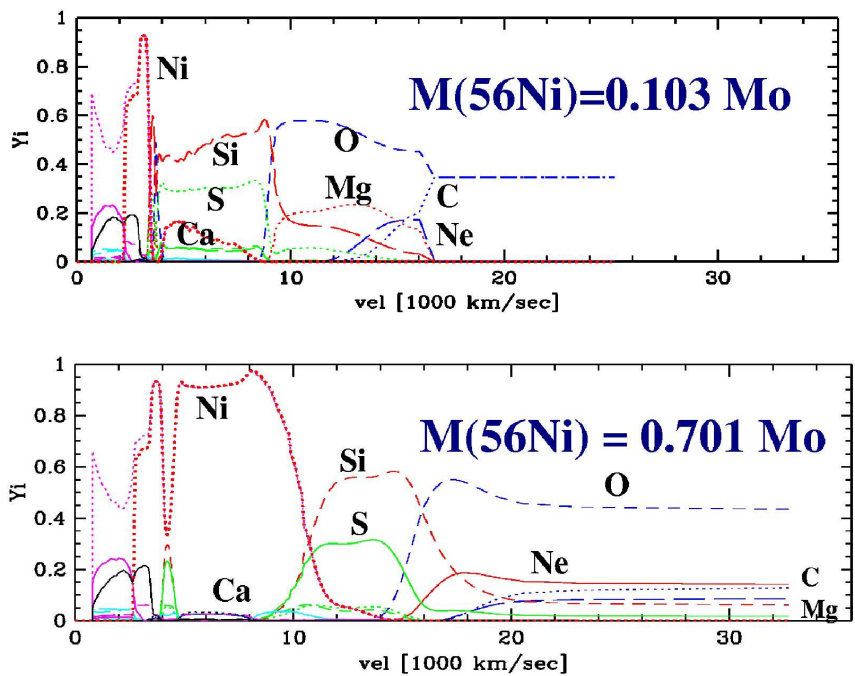}
\caption {Final chemical structures of models are given for  
a strongly subluminous (top panel) and a normally-bright (bottom panel)
supernova during the phase of homologous expansion.
 Identical progenitors and prescription for the deflagration 
and detonation fronts have been used but the deflagration-detonation
transition density, $\rho_{tr}, $ of $1 \times 10^7$ g~cm$^{-3}$ and 
$2.5 \times 10^7$ g~cm$^{-3}$ have been used to produce the subluminous 
and normally-bright models, respectively.  Note that in the subluminous
model, \Ni56\ is confined to a lower region of velocity space, beneath the
optical photosphere at maximum light.}
\label{model1}
\end{figure}

\begin{figure}[th]
\includegraphics[width=9.5cm,angle=270]{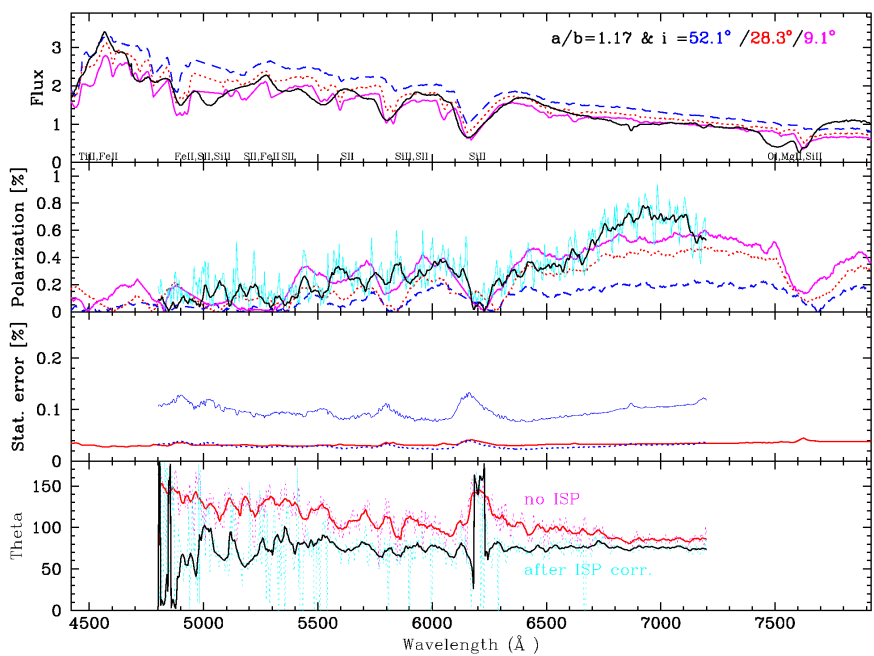}
\caption {Flux spectra (top panel) and polarization spectra 
(second panel) at day 15 after the explosion
for a subluminous delayed-detonation model (see text). 
The observations are not corrected for reddening, since it
is uncertain, though they are corrected for the negligible 
redshift ($z=0.00213$).
The model SN has been mapped into an oblate ellipsoid with axis ratio 
1.17 shown for various inclinations $i$ and compared to the observations  
of SN 1999by at about maximum light in B (see \S 2).
For comparison with the models, an interstellar
component with $P=0.25\%$ and $\Theta = 140\deg$ has been 
subtracted from the observations.
The observed flux spectrum is the solid black line.
Model flux spectra are shown by solid magenta, dotted red, and dashed
blue lines for inclination angles, $i$, of 9.1\deg, 28.3\deg, and 
52.1\deg, respectively.  In the second panel, the raw polarization data is 
shown by a light blue line, while the smoothed data (with $\Delta 
\lambda = 45$\AA) is the black line.  The model polarization spectra are 
the same as the first panel.  The third panel presents the statistical error 
in $P$ 
for the raw data (blue), smoothed data (dotted blue), and theoretical 
model (red; binned to $\Delta \lambda = 12.5 $\AA). 
The model seen nearly equator-on ($i=9.1\deg$) does the best job of
reproducing the general features of the polarization spectrum:  
depolarization from 4900 \AA\ to 5500 \AA, moderate polarization 
from 5500 \AA\ to 6100 \AA, a depolarization in the Si 6150 \AA\ feature,
and a rising polarization to the red.
In the bottom panel, the observed polarization angle $\Theta $ is 
shown as a function of wavelength. The unbinned data are given by
dashed lines and the binned data by solid lines.  The data are
shown before and after correcting for the ISP contribution determined 
from the models (see text).  
After correction for ISP is clear that the position angle is 
essentially constant across all wavelengths.  There
is an $80\deg$ shift across the \SiII\ trough feature, though
this is an artifact of low polarization close to the origin giving
an uncertain position angle.  Compare this choice of
ISP to the empirically determined choice, Figure \ref{quptbestisp}.}
\label{model2}
\end{figure}

\begin{figure}[th]
\includegraphics[width=9.5cm,angle=270]{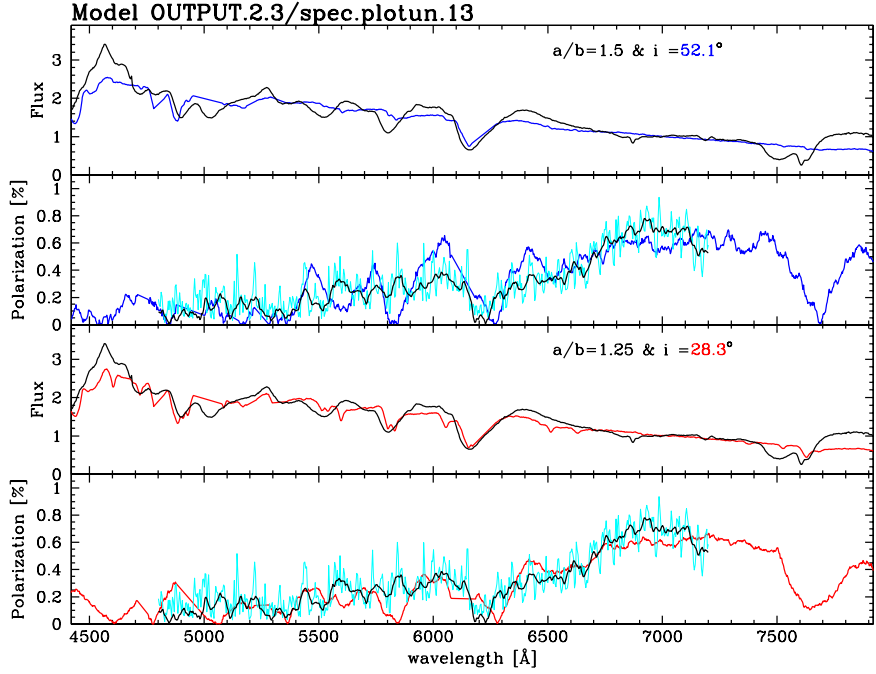}
\caption {Same as Figure \ref{model2} but with $a/b=1.5 $ and  
$ i = 52.1\deg$ (upper two panels) and with $a/b=1.25$ and 
$ i = 28.3\deg$ (lower two panels) in comparison with the observations  
of SN 1999by. These models, particularly the model with $i=52.1\deg$ do 
not match the data as well as the nearly edge-on geometry of Figure \ref{model2}.}
\label{model3}
\end{figure}

\begin{figure}[!htp]
\begin{center}
\plotone{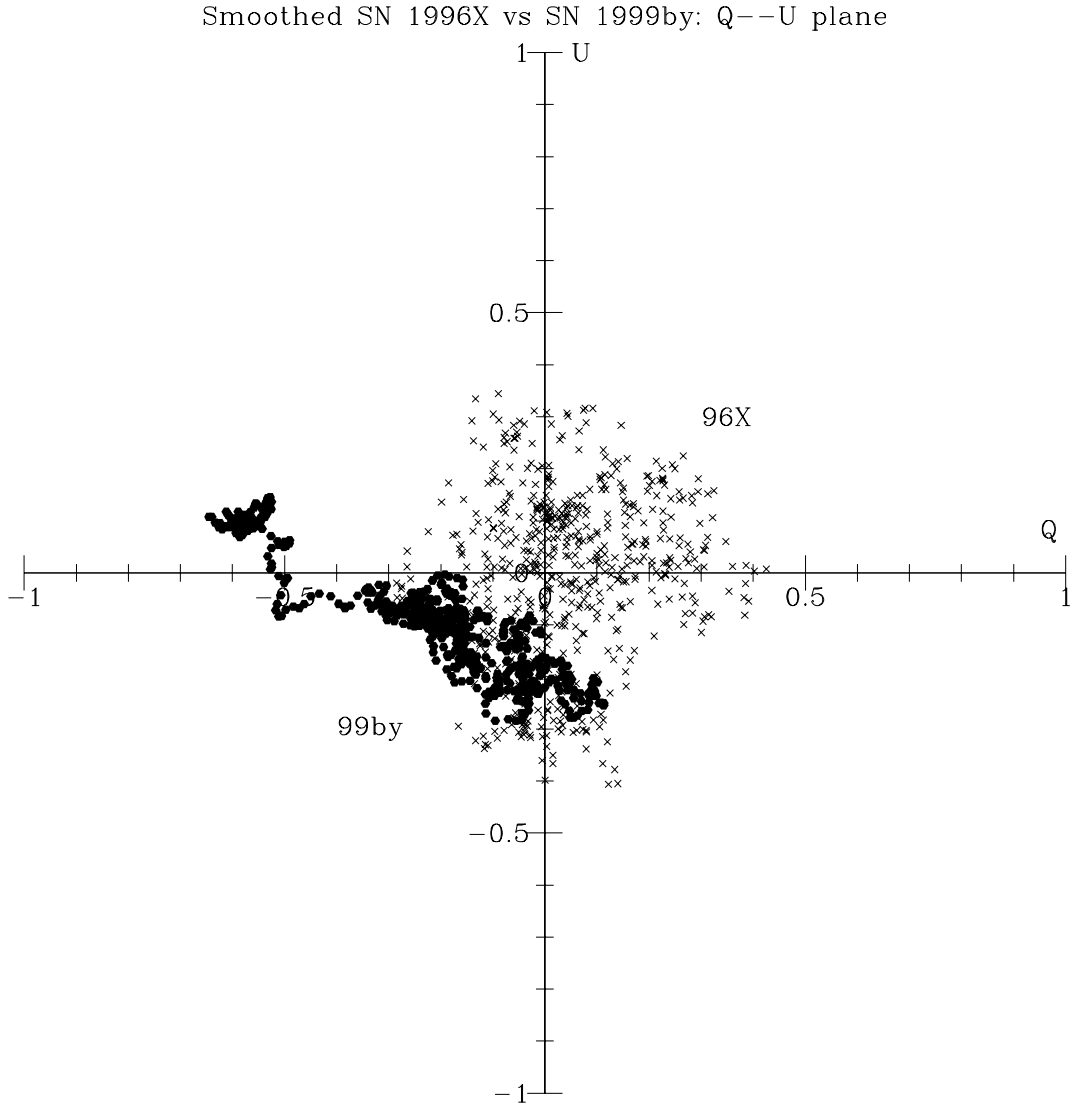}
\end{center}
\caption[All observations:  Smoothed, \qu\ plane for SN 1999by vs SN 1996X.]{Smoothed,  \qu\ plane for SN 1999by vs SN 1996X created by summing all observations.  Red crosses are SN 1996X and blue solid points are SN 1999by.}  
\label{qucomp}
\end{figure}

WWH97 observed the normally-bright supernova
SN~1996X about one week before maximum light in V.
The data showed a small average polarization with
modulations on the level of 0.2\%. 
WWH97 concluded that SN~1996X could have only a very small,
if any, global asymmetry in the geometry, given the lack of
detectable mean polarization, but that the modulations
were consistent with inhomogeneities in distributions of the elements 
within the ejecta.  Figure \ref{qucomp} shows the data for SN~1996X and
SN~1999by in the \qu\ plane.  The lack of any preferred orientation to
the data and hence a global asymmetry for SN~1996X contrasts distinctly
with the linear distribution of points for SN~1999by.  The finer
scale variations in the polarization spectra of SN~1996X are
qualitatively similar to the expectations for the behavior of
a photosphere significantly contaminated with iron-peak elements
as a normally-bright SN~Ia should be (Figure \ref{model1}, WWH97).
A comparison with detailed model
calculations for a delayed detonation model confirmed that the
data are consistent with little or no global 
asymmetry in the density structure of SN~1996X. 
The continuum polarization in the model is modulated by
ubiquitous weak lines of Fe~II-III, Co~II-III and Ni~II-III  
that serve to depolarize on scales of $\approx 100$ \AA. 
These lines are formed at the interface region
between complete and incomplete Si burning. 
WWH97 suggest that there is an asymmetry in the {\it chemical} structure 
that may be a relict 
of the chemical plumes rising during the deflagration phase.
Strong depolarization over the \SiII\ 6150 \AA\ feature is
predicted by the model, but not observed.
As noted in WWH97, another problem with the comparison to theory of SN~1996X
is that the model shows a rise in the polarization longward of
6500 \AA\  which is not apparent in the data.  SN~1999by shows both
the depolarization at the \SiII\ feature and the rise in the polarization
level to the red.  

One of the open questions in SNe~Ia research is the 
nature of the very subluminous subclass.
The prototypical example of this subclass is SN~1991bg.  
Light curves and spectra of SN 1991bg are presented
in F92, Leibundgut \etal\ (1993), and Turatto \etal\ (1996).
Other subluminous SNe include SN 1992K (Hamuy \etal\ 1994), SN 1997cn 
(Turatto \etal\ 1998) and SN 1998de (Modjaz \etal\ 2000).
Subluminous SNe~Ia may be rare, but some
analyses indicate that they could make up 16\% (Li \etal\ 2000) 
or more (Schaefer 1996) of all local SNe~Ia.
Some defining characteristics of the subclass are
rapidly-declining light curves ($\drp \simeq 1.9$),
fainter than normal peak magnitudes by 2--3 mag,
redder colors at maximum ($(B-V)_{\rm max}$ $\simeq$ 0.4--0.74),
a non-existent or weak secondary maximum in R and I, and
an earlier  transition to the nebular phase. Moreover,
the \Ni56\ region is limited to expansion velocities below
4000 to 4500 km~s$^{-1}$, i.e. about a factor of 
2 to 3 times smaller than in normally-bright SNe~Ia.  By comparison,
the expansion velocity implied  by Si II  
indicates a wider range of the Si-dominated layers in subluminous SNe~Ia 
($\approx 5000 $ to 12,000  km~s$^{-1}$)
compared to normally-bright SNe~Ia ($\approx 9000 $ to 13,000  km~s$^{-1}$;
Turatto et al. 1996; Gerardy et al. 2000).
Subluminous events also show strong absorption features at 
4200 \AA\ (\TiII), 5800 \AA\ (\SiII), 7500 \AA\ (\OI\ \& \MgII) 
and 8300 \AA\ (Ca II).
Theoretical interpretations of subluminous SNe~Ia given 
in Ruiz-Lapuente \etal\ (1993), Woosley \& Weaver (1994), HKW95, 
N95, Mazzali \etal\ (1997), and Milne \etal\ (1999) include
all three types of explosion mechanisms outlined in \S 1.

SN~1999by is the first subluminous SN~Ia to 
show definitive evidence for intrinsic polarization.  Before ISP correction,
the supernova shows an overall level of polarizaion, a rise in $P$ redward
of 6500 \AA, and a change in
polarization position angle across the \SiII\ 6150 \AA\ feature. The            
strong wavelength dependence and the individual
features are not expected from interstellar
polarization and show the need for a component intrinsic to the supernova.
 
We employ a new method (see Wang et al. 2000b)  
for the  analyses the polarization data of supernovae by using the \qu\ plane.
This representation provides important constraints on the overall 
geometry of the configuration, and it allows strong constraints to be
placed on the ISP if the data points are aligned.
The small spread of the data points in SN 1999by about a line
in the \qu\ plane reveals that the SN has a well-defined axis of
symmetry. Though the spread around this line is consistent 
with rotational symmetry, we cannot exclude small additional 
off-axis components. 
In the classical approach to determine the polarization component of the ISM,
 the object needs to be observed at several epochs from
which the polarization vector can be deduced by the reasonable assumption that it is time invariant.
 In the \qu\ plane and for reasonable assumptions for the polarization mechanism, we have shown
that the component of the ISM can be determined by the vector between the origin of the
\qu\ plane and the endpoint of the line of the polarization data
points. Obviously,
with this choice, the  polarization position angle should not vary over lines. 
The residuals can be used to judge the quality of the data and the assumptions.
 For SN 1999by, this method suggests that the $P_{\rm ISP}$ $=0.3\%$ with
$\Theta=150\deg$ which is consistent with the best fits for the ISM based on detailed models
($P_{\rm ISP}$$=0.25 \%, \Theta=140\deg$).
 
 Based on a delayed detonation model, synthetic NLTE-spectra 
have been computed and compared to the observations. 
The free parameters have been adopted from a spherical 
delayed detonation model that reproduces the infrared spectrum 
of SN~1999by between 1 and 2 $\mu m$ and its evolution with time 
(Gerardy et al. 2000). No further tuning of the initial model 
is done to fit the polarization data, although three new parameters
must be introduced, the axis ratio, the angle of the line of sight
to an observer, and the ISP.  The ISP could be taken from the
observations, but the theoretical analysis independently confirms the empirical
value.  The axis ratio and inclination angle are constrained
by two aspects of the observations: the overall level of 
polarization and the shapes of spectral features.   
The initial spherical model is remapped to an 
ellipsoidal geometry with an axis ratio of $a/b$.
The axis ratio $a/b$ has been adjusted to provide
the overall level of $P$ observed.
Both the optical flux spectra and polarization spectra of 
SN~1999by can be reproduced by models with an
asphericity of $\approx 20 \% $ which is seen equator on. 
Higher inclinations, which require larger axis ratios to give
the overall polarization, can be ruled out
from the spurious frequency dependence of the resulting
polarization spectra.

The relatively high polarization we observe for SN~1999by
may be a significant clue to the nature of SNe~Ia.
Existing limits on polarization of other SNe~Ia are very
sparse (7 objects at the time of this paper, see Table 2).
The polarization due to asphericity  
decreases with inclination, further reducing the significance
of any statistics.  Within these limitations,
we would have expected to see more polarized SNe~Ia if 
$P \approx 0.6$ to $0.8\% $ is common.  The lack of such detections  
supports the notion that the intrinsic polarization of the subluminous
SN 1999by is unusually large.
Another argument is that, although small, the size of the asymmetry we derive
implies a change of the observed luminosity
of about $0.3^m$ from equator to pole.
This spread is larger by a factor of $\approx 1.5 $
than the  mean dispersion in the brightness-decline relation $M(\drp )$
(Hamuy et al. 1995; Riess et al. 1999).
The dispersion in $M(\drp )$ is dominated by the normally-bright SNe~Ia and
the orientation of the symmetry axis of a given supernova is arbitrary. 
Taken together, these facts again suggest that the asphericity is 
unusually large for SN 1999by compared to most normally-bright SNe~Ia.
The low mean polarization observed for the normal SN~1996X (WWH97) is
consistent with this.  
If future observations confirm that the large asphericity 
of SN~1999by is characteristic of subluminous SNe~Ia, 
this may provide important clues 
to the physical reasons for both normal and less-than-normal 
peak luminosity.

Our quantitative analysis of the data of SN 1999by was based on 
delayed-detonation models with imposed asphericity without addressing 
the question of the physical mechanism which produces aspherical envelopes. 
Within the delayed detonation model, possible mechanisms to 
induce asymmetry include the following:
1) instabilities in the nuclear burning front during the deflagration phase,
2) rapid rotation of the progenitor white dwarf,
3) delayed detonation transition at a point rather than 
simultaneously on a sphere, and 
4) impact of the supernova ejecta on the secondary star.
Aspherical configurations may also be the result of 
5) the merging process of two degenerate white dwarfs and 
6), in principle, edge-lit helium detonations in sub-Chandrasekhar mass WDs.

Studies of the deflagration phase of DD models reveal large
plumes of burning material 
(e.g. Khokhlov 1995, 2000; Niemeyer \& Hillebrandt 1995). 
These plumes may leave their imprint on the chemical structure of the ejecta. 
The plumes do not significantly perturb the density distribution and hence
will not explain global asymmetries in the overall density structure 
(Khokhlov 2000), but could leave an imprint at composition interfaces. 
WWH97 suggested this mechanism as a possible
explanation for the polarization signatures  seen in SN~1996X, but 
these plumes are an unlikely explanation for the  
global deviations from sphericity with a well-defined 
axis of symmetry as in SN~1999by.

 Livne (1999) pointed out that we cannot expect a deflagration/detonation
transition on a sphere as implicitly assumed in spherical models, 
but that the transition may begin
at one point. Detailed 2D calculations showed large scale
asymmetries, but Livne only followed the explosion process. He
did not follow the calculation into the homologous expansion, which
may or may not destroy any initial asphericities 
by the time of maximum light when the observations of SN~1999by were taken.  
It is also not clear why such a mechanism should apply to subluminous
SNe~Ia and not, perhaps, to most SNe~Ia.

 In any accretion model, we  expect an impact of the supernova ejecta with the
secondary star.  Marietta, Burrows, \& Fryxell (2000) computed 2D numerical 
simulations of the impact of an SN~Ia explosion with hydrogen-rich main 
sequence, subgiant, and red giant companions.  They find that the blast 
strips main sequence and subgiant companions of 15\% of their mass while 
red giants lose $\sim$ 97\% of their envelopes (0.5 \Msun ).
The impact of the ejecta with the secondary star creates a hole in the
debris of angular size $\sim 30\deg-40\deg$, corresponding to $\sim$ 10\%
of the ejecta surface.  The result was similar for all
cases considered because the companion was close enough to be in 
Roche lobe overflow. This effect may be consistent 
with the polarization observed in SN 1999by,
but we would expect the same effect in both normal and subluminous 
SNe~Ia which may be in conflict
with the low polarization observed in normally-bright SNe~Ia.

Mahaffey \& Hansen (1975) calculated a rotating detonation model,
but no detailed models have been calculated for rapidly rotating, 
deflagrating white dwarfs. Rapid rotation will produce  
a global distortion of the density structure of the presupernova WD 
(M\"uller \& Eriguchi 1985) and may also affect the propagation of 
nuclear burning fronts in the supernova, resulting in small asphericities. 
We regard this possibility as very attractive since the combined
effect of rotational distortion and modification of the nuclear 
deflagration front has the potential to link the subluminosity 
of SN 1999by to the large, global asymmetry.
A comparison with Model 7 of M\"uller \& Eriguchi suggests that 
even with solid body rotation (which produces the minimum
distortion for a given angular momentum), sufficient rotation
to distort the structure in the silicon layers of our subluminous
model would need to have a rotation energy only about 5\% of
the gravitational energy.  Thus a significant, but not extreme, 
rotation may be sufficient to explain the degree of
polarization we observe in SN~1999by. 
Rotation could thus serve as a single parameter that distinguishes
normal SNe~Ia arising in slowly-rotating white dwarf progenitors 
from subluminous SNe~Ia that occur in more rapidly-rotating white dwarfs.
One possible problem with the scenario of rotation as a single parameter
is that SN properties appear to show a correlation with environment
and by implication, progenitor age (Howell 2000a,b).

In the merging scenario, two C/O white dwarfs merge to produce an SN Ia
(Webbink 1984; Iben \& Tutukov 1984).  This would provide a natural
explanation for the lack of hydrogen seen in Ia spectra.  There is
evidence that such systems exist (Saffer et al.\ 1998).  Khokhlov et al. 
(1993) and HK96 parameterized merger events as spherically symmetric 
C/O WDs with thick envelopes.  They produced light curves that were in 
reasonable agreement with some SNe~Ia.  Doubts about the merger models 
have been raised.  Three-dimensional hydrodynamical simulations of the 
process show that (unless the system ignites during the merger process)
the less massive star is disrupted into a hot envelope 
and an accretion disk around the primary (Benz \etal\ 1990; Rasio \& 
Shapiro 1995; Mochkovitch, Guerrero, \& Segretain 1997).  
In this case, off center carbon ignition produces a flame that 
propagates inward, converting the star to an O+Ne+Mg WD (Nomoto \& Iben 1985;
Saio \& Nomoto 1985, Kawai, Saio, \& Nomoto 1987; 
Saio \& Nomoto 1998).  Magnesium-24 then undergoes electron capture, 
inducing accretion-induced collapse, to produce a neutron star
(Saio \& Nomoto 1985, Mochkovitch \& Livio 1990, Nomoto \& Kondo 1991).
Alternatively, it may be possible to ignite the system due to tidal heating
during the merger process (Iben, Tutukov, \& Fedorova 1998) which could
produce a highly aspherical explosion.  This scenario has not been
demonstrated in realistic simulations, but deserves further study.

We have to stress the limits of this study which
must be seen only as  a first step.
Clearly, more high quality data must be obtained to provide 
a statistical sample and to make use of small scale features 
to evaluate details of the geometry and, ultimatively,
to probe details of the explosion mechanism (e.g. the burning properties) 
and the scenario (e.g. rotation of a single rapidly rotating WD vs. 
merging of two WDs).  Future data should have a noise level well 
below 0.1\% to address questions of
deviations from axial symmetry and to utilize the small 
scale variations seen in the models.  This is a very feasible goal with  
the current and upcoming generation of 8 m class telescopes.
Our models are based on aspherical envelopes which are distorted artificially. 
Although this is a reasonable approach to estimate the size of 
asphericity and general properties, it hardly
provides the desired link to the physical mechanisms responsible. 
Detailed hydro calculations for the various mechanism are needed.
Within the delayed detonation scenario, consistent hydro calculations 
of rotating WDs, deflagration fronts and the
multi-dimensionality of the deflagration to detonation transition are feasible.
Although any successful model must reproduce the overall pattern in the 
chemical composition, alternative scenarios with full multi-dimensional
hydrodynamics may do this job as well. 
In particular, the merging scenario needs more attention.

\acknowledgements
This research was supported in part by NSF Grant 95-28110,
by  NASA Grant LSTA-98-022 and a grant from the Texas Advanced Research Program.
 The calculations  were done on a cluster of workstations 
financed by the John W. Cox Fund  of the Department of Astronomy at the
University of Texas, and processors donated by Advanced Micro Devices (AMD). 
We are grateful to A. Filippenko, D. Leonard, and an anonymous referee 
for discussions and to G. Hill for providing the IGP which made this work 
possible.

\noindent{\bf APPENDIX: Computational Methods}

\noindent
{\bf 1. Radiation-Hydrodynamics and Light Curves in Spherical Geometry}

  Explosion models are  calculated using a one-dimensional
radiation-hydro code (HK96)  that solves the hydrodynamical
equations explicitly by the piecewise parabolic method
(Colella \& Woodward 1984).
 Nuclear burning is taken into account using an extended network of 216
isotopes (Thielemann, Nomoto \&   Hashimoto 1996 and references therein).
 The propagation of the nuclear burning front is given by the
velocity of sound behind the burning front in the case of a detonation 
wave and in a parameterized form during the deflagration phase based
on detailed 3-D calculations (e.g.  Khokhlov 1995, 2000; Niemeyer \& Hillebrandt 1995).
  We use the parameterization as described in Dominguez \& H\"oflich (2000).
 We assume that $v_{\rm burn}=max(v_{t}, v_{l})$,
where $v_{l}$ and  $v_{t}$ are the laminar and turbulent velocities with
$$  v_{t}= C_1~\sqrt{\alpha_{T} ~ g~L_f},~~~ ~\eqno{[A1]} $$
with $ C_1=0.15$ and with
$$\alpha_T ={(\alpha-1)/( \alpha +1}), \eqno{[A2]} $$
and with 
$$\alpha ={\rho^+(r_{\rm burn})/ \rho^-(r_{\rm burn})}.  \eqno{[A3]} $$
\noindent
Here $\alpha _T$ is the Atwood number, $L_f$ is the characteristic
length scale, and $\rho^+$ and $\rho^-$ are the densities in front of
and behind the burning front, respectively. The quantities
 $\alpha$ and  $L_f$  are directly taken
from the hydro at the location of the burning front
and for this choice
of $C_1$ we take $L_f=r_{\rm burn(t)}$.
The transition density is treated as a free parameter.
The description of the
deflagration front does not significantly  influence
 the final structure of the explosion.  The amount of matter consumed
during burning (and the total $^{56}$Ni production) is governed by
 the pre-expansion of the WD
and, consequently, is determined by the transition density $\rho_{tr}$,
at which the burning front switches from
the deflagration to the detonation mode (H\"oflich 1995a).
The  value $\rho_{tr}$ can be adjusted to produce a given amount of \Ni56\ which
determines  the brightness (H\"oflich 1995a) and through the temperature
dependence of the opacity, the decline rate (HKW95; H\"oflich et al. 1996).
 
The code also simultaneously solves for the energy and  radiation transport with variable
Eddington factors.
The radiation transfer portion of the code
consists of (i) an LTE radiation transfer scheme 
based on the time-dependent moment equations which are
solved implicitly, (ii) a frequency-dependent radiation transport to determine the
Eddington factors and the frequency averaged opacities,
 (iii) a detailed equation of state with an elaborate
treatment of the ionization balance and the ionization energies, (iv)
time-dependent expansion opacities  which take into account the
composition structure of the explosion model, (v) photon scattering and thermalization calibrated
by NLTE-calculations, and (vi) a Monte Carlo
$\gamma$-ray deposition scheme which takes into account all relevant
$\gamma$-ray transitions and interaction processes. For more details, see H\"oflich et al. (1999), and references therein.

\noindent
 {\bf 2. Density and Chemical Structure of the 3-D envelope}

 Aspherical density structures are constructed from the
spherical density distribution by
imposing a  homologous expansion function which depends on the angle $\Theta$
from the equatorial plane and which conserves the total energy and
the mass fraction per steradian from the spherical model (H\"oflich et al. 1999).
The spherical density structure is given by
$$\rho(R, \Theta ) = \rho (R), \eqno{[A4]}$$
\noindent
with R the initial distance of a mass element from the center.
Homologous expansion of this density distribution is assumed
with a scaling that depends on angle, i.e., 
$${v(R,\Theta) \over R} = C(\Theta),\eqno{[A5]}$$
\noindent
so that
$$r(\Theta ) =  C(\Theta ) ~R ~t,\eqno{[A6]}$$
\noindent
where $t$ is the time since the explosion
and $ r(\Theta)$ is the distance of the 
mass element after time $t$. 
 
Because little is known about the general geometry of the ejecta,
we construct  ellipsoidal isodensity contours with an axis ratio
$E= a/b$ at the photosphere where
 $a$ is the distance of the photosphere  in the $x-y$ (symmetry) plane and
$b$ is the distance in the (axial) $z$-direction.          
This  contour is given by
$$ r(\Theta) = r(\Theta = 0)~{\sqrt {(\cos^2(\Theta ) + 
    E^2 ~ \sin ^2(\Theta ))}}.\eqno{[A7]} $$     
The homology scaling constant, $C(\Theta)$,  is
determined to produce the desired axis ratio for an ellipsoid (H\"oflich et al. 1999),
and the total explosion energy is normalized to that of the spherical model.
Here, the asphericities are small and the
temperature profiles in SNe~Ia are shallow (HMK93,
H\"oflich 1995a). Therefore, we assume identical isodensity 
and isotemperature contours.

\noindent
{\bf 3. Spectra for Asymmetric Configurations}

 Our radiation transport code  works both  for spherical geometry and 
in three dimensions using different modules for the radiation transport.
 For  details and references to  the atomic database,  
 see H\"oflich (1995a) and references therein.
 The density and abundance structure is taken from the hydro calculations.
Excitation due to $\gamma$-rays is included via the Monte
Carlo code.
 Based on the list of atomic data of Kurucz (1995), we have constructed
detailed atomic models for a couple of ions similar to those used in  WWH97.
 The use of 3-D geometries caused some restrictions on the complexity of the atomic
models. Multilevel atoms have been restricted to the main ionization species of
\ion{C}{1} (23/57), \OI\ (28/75), \MgII\ (25/85), \SiII\ (28/83), \ion{Ca}{2} (26/90) of  the intermediate mass
elements. Here, the numbers in the parentheses denote the number of levels and 
of  line transitions, respectively. For those ions,
 the statistical equations are solved 
consistently including bound-bound and bound-free transitions.
  The method of  accelerated lambda iteration is applied to remove the global
dependence of the level populations  and the radiation field. 
 In order to include the line blocking effect, about 1,000,000 
additional lines are taken  into account under the assumption of local thermodynamical equilibrium
for the population number but including scattering terms calibrated by the NLTE-elements.
 The electron temperature structure is based on the depth-dependent
luminosity from the monochromatic light curve calculation at a certain time.
The complete system of equations is given by
 the time-independent radiation transport  and
 the statistical equations, i.e.,
 the rate equation,
 the particle conservation equation, and
 the charge conservation equation.
All allowed bound-bound and bound-free transitions between the NLTE levels
are taken into account. Complete redistribution over each individual
line both in frequency and in angle are assumed (Mihalas 
1978) in the comoving frame (but not the observer's frame).
 This means that the populations
of sublevels of the upper and lower transition are described by
a Maxwell-Boltzman distribution and that the light becomes
unpolarized.
 
\noindent
{\bf 3.1 Radiation Transport in 3-D  NLTE-Spectra}
 
The current calculations use a modified 
version of the Monte-Carlo code previously applied to 
calculate the continuum in SN 1987A (H91) and line polarization in SN 1993J 
and SN 1996X (H\"oflich et al. 1995b; WWH97).
 The code is capable of handling arbitrary
3-dimensional geometries, both for the density and the distribution of the 
sources. 
 Polarization and flux spectra for rapidly expanding envelopes can 
thus be computed.
Polarization is treated within the Stokes formalism (see, e.g., Van de Hulst, 1957).
We include electron scattering as a polarization process but
omit scattering at dust particles.                                                    

 The modifications for the present work
to the previous versions of our Monte Carlo scheme (H91)
  are mainly related to the inclusion of NLTE-effects. In
scattering dominated atmospheres, deviations from  local thermodynamic equilibrium
are also relevant at large optical depths; however,
 at large optical depths, Monte-Carlo methods become very costly and/or inaccurate
(H\"oflich et al. 1995b). Therefore,
 we use a hybrid scheme of Monte-Carlo and non-equilibrium diffusion methods.
 In the latter case,
 we implicitly solve the time dependent radiation transport equation
in a non-equilibrium, diffusion approximation for 3-D geometry including the
scattering and thermalization terms for the source functions and include the 
frequency derivatives in the formulation for the opacities and emissivities
(Lucy \& Solomon 1970; Karp et al. 1977).
  At large optical depths, this provides the solution for the
full radiation transport problem.
 We note that the use of a non-equilibrium diffusion does {\it not} 
imply that the mean intensity J is given by the black body field.
 To obtain the correction 
 solution for the radiation transport equation at small optical depths, the difference
between the solution of the diffusion and full radiation transport equation is calculated
by a Monte Carlo method. We calculate the difference between the solutions for computational
accuracy and efficiency.
 Consistency between the solution at the outer and inner region
is obtained  iteratively.
The same Monte Carlo solver is  used which has been applied  to compute $\gamma$-ray
and  polarization spectra (e.g. H\"oflich et al. 1995b).
 The Monte Carlo method is appropriate for this problem because of its flexibility 
with respect to the geometrical and velocity structures.

\noindent
{\bf 3.2 Coupling of the Radiation Transport and Rate Equations}
 
 The solution of the radiation transport and the statistical equations for the
level populations are coupled. A perturbation method is used to obtain 
consistency between the solutions (H\"oflich 1995a).
 In the equations for the microphysical quantities
 we express the actual mean intensity J$_{\nu}$ by the following
equation:
 
$$ J_{\nu}^{(m)}= \Lambda_{\nu}^{(m)} S_{\nu}^{(m)} \approx
\Lambda_{\nu}^{(m-1)} S_{\nu}^{(m-1)} + \Lambda_{\nu}^{* ~(m-1)}
(S_{\nu}^{(m)}-S_{\nu}^{(m-1)}). \eqno{[A8]}$$
 
\noindent 
Here, the indices in brackets denote the iteration step, and
$S_{\nu}$ and $\Lambda_{\nu}$ are the source function and the
radiation transport matrices, respectively. 
The matrix $\Lambda^*$ is a band-matrix with elements corresponding
to the diagonal and first off-diagonal elements of the complete matrix $\Lambda$.
The elements of $\Lambda $ are computed 
in the narrow line limit (Sobolev et al. 1957; H\"oflich 1995a).

 The use of a Monte-Carlo scheme introduces an additional complication due
to the ``photon" noise.
 To achieve numerical stability for the solution in the rate equations,
the deviation from the solution for $ J_\nu$
 obtained by the Monte-Carlo calculations
 and the non-equilibrium diffusion are set to
zero if they are less than the statistical noise.  This avoids instabilities 
during the model iterations when using the accelerated lambda iteration technique;
 however, this also implies an increase of the computational expense with increasing
rate of convergence.  To increase the photon statistics, solutions of the 
Monte-Carlo calculation from previous
iterations, $n$, are taken into account with individual weight factors. 
Currently, we use a $1/(m-n+1)$ weight where $m$ is the total number of model 
iterations. Clearly, this weight function needs further fine tuning.
 Tests for realistic structures of the envelope show that the resulting error in the population
numbers inherent to our procedure is comparable to those introduced by discretization errors
if we solve the radiation transport equation in a difference scheme.

\newpage
\clearpage

\newpage
\clearpage

\newpage
\clearpage

\newpage
\clearpage

\newpage
\clearpage

\newpage
\clearpage

\newpage
\clearpage

\newpage
\clearpage

\newpage
\clearpage
 
\newpage
\clearpage

\newpage
\clearpage
 
\newpage
\clearpage

\newpage
\clearpage

\begin{deluxetable}{lrrrrlrr}
\tablewidth{0pt}
\tablecaption{Observing Log\label{log99by}}
\tablehead{
\colhead{Object}& 
\colhead{Exptime\tablenotemark{a}}&
\colhead{UT Start\tablenotemark{b}} & 
\colhead{UT Date} & 
\colhead{PA\tablenotemark{c}} &
\colhead{Type\tablenotemark{d}}& 
\colhead{Par\tablenotemark{e}}& 
\colhead{X\tablenotemark{f}} \\
&\colhead{(s)} &\colhead{(h:m)} &\colhead{(m/d/y)}&\colhead{(\deg )}&&\colhead{(\deg )} }
\startdata
SN1999by&1200&4:25&5/8/99&0&o&11.8&1.31\nl
SN1999by&1200&4:47&5/8/99&45&o&6.9&1.38\nl
SN1999by&1200&5:10&5/8/99&22.5&o&2.1&1.47\nl     
SN1999by&1200&5:32&5/8/99&67.5&o&-2.2&1.58\nl     
SN1999by&1200&6:03&5/8/99&0&o&-7.9&1.77\nl       
SN1999by&1200&6:25&5/8/99&45&o&-7.3&1.94\nl           
SN1999by&1200&6:47&5/8/99&22.5&o&-15.6&2.16\nl            
SN1999by&1200&7:08&5/8/99&67.5&o&-19.2&2.42\nl         
HD 154445&3&9:55&5/8/99&0&p&68.1&1.21\nl
HD 154445&3&9:56&5/8/99&45&p&67.8&1.21\nl
HD 154445&3&9:57&5/8/99&22.5&p&67.4&1.21\nl
HD 154445&3&9:58&5/8/99&67.5&p&67.1&1.22\nl
HD 192281&25&10:26&5/8/99&0&f&17.7&1.08\nl
HD 245310&240&3:02&5/9/99&0&p&-27.4&3.70\nl
HD 245310&240&3:07&5/9/99&45&p&-27.7&3.96\nl
HD 245310&240&3:14&5/9/99&22.5&p&-28.1&4.39\nl
HD 245310&240&3:19&5/9/99&67.6&p&-28.4&4.76\nl
SN 1999by&1800&4:22&5/9/99&0&o&11.6&1.31\nl
SN 1999by&1800&4:58&5/9/99&45&o&3.7&1.44\nl
SN 1999by&1800&5:31&5/9/99&22.5&o&-2.7&1.60\nl
SN 1999by&1800&6:04&5/9/99&67.5&o&-8.8&1.81\nl
SN 1999by&1800&6:40&5/9/99&0&o&-15.1&2.13\nl
SN 1999by&1800&6:56&5/9/99&45&o&-17.9&2.31\nl
SN 1999by&1800&7:12&5/9/99&22.5&o&-20.6&2.53\nl
SN 1999by&1800&7:28&5/9/99&67.5&o&-23.3&2.80\nl
HD 155528&120&8:49 &5/9/99&0&p&85.3&1.22\nl
HD 155528&120&8:52 &5/9/99&22.5&p&86.5&1.22\nl
HD 155528&120&8:55 &5/9/99&45&p&87.6&1.22\nl
HD 155528&120&8:59 &5/9/99&67.5&p&89.1&1.22\nl
BD +40 4032&240 &9:42&5/9/99&0&f&10.3&1.14\nl
SN1999by&1800&3:07&5/10/99&0&o&32.6&1.15\nl
SN1999by&1800&3:39&5/10/99&45&o&22.6&1.20\nl
SN1999by&1800&4:12&5/10/99&22.5&o&13.9&1.28\nl
SN1999by&1800&4:46&5/10/99&67.5&o&6.2&1.39\nl
SN1999by&1200&5:21&5/10/99&0&o&-0.8&1.54\nl
SN1999by&1200&5:42&5/10/99&45&o&-4.8&1.66\nl
SN1999by&1200&6:03&5/10/99&22.5&o&-8.6&1.80\nl
SN1999by&1200&6:24&5/10/99&67.5&o&-12.3&1.97\nl
BD +22 3782&180&8:30&5/10/99&0&p&20.5&1.31\nl
BD +22 3782&180&8:35&5/10/99&45&p&20.5&1.29\nl
BD +22 3782&180&8:39&5/10/99&22.5&p&20.5&1.27\nl
BD +22 3782&180&8:43&5/10/99&67.5&p&20.6&1.26\nl
BD +25 3941&240&9:01&5/10/99&0&f&26.6&1.19\nl
\enddata
\tablenotetext{a}{Exposure time}
\tablenotetext{b}{UT at Observation Start}
\tablenotetext{c}{Waveplate Position Angle}
\tablenotetext{d}{Observation type: o=object (SN), p=polariz. std., f=flux std.}
\tablenotetext{e}{Slit orientation with respect to Parallactic Angle}
\tablenotetext{f}{Airmass}
\tablenotetext{}{
\,\newline
May 8 Weather Note --- Clear.  Moonrise 7:26 UT, after SN obs.\\
May 9 Weather Note --- Some clouds scattered at beginning of night, dissipated.  Moonrise 8:08 UT, after SN obs. \\
May 10 Weather Note --- Some clouds scattered at beginning of night, quickly
dissipated.  Wind picked up towards the end of the night.  Moonrise 8:49 UT, 
after SN obs.\\}
\end{deluxetable}

\begin{deluxetable}{llcrrlcc}
\tablewidth{0pt}
\tablecaption{SNe Ia with polarization measurements\label{polSNe-table}}
\tablehead{
\colhead{SN}& 
\colhead{Galaxy}&
\colhead{Spec\tablenotemark{a}}&
\colhead{Sub\tablenotemark{b}}&
\colhead{Epoch\tablenotemark{c}}& 
\colhead{p (\%)}&
\colhead{Detection\tablenotemark{d}}&
\colhead{Ref.}
}
\startdata
1972E&NGC 5253&b&n&30:&0.35$\pm$0.2&Undetermined&1\nl
1975N&NGC 7723&b&n:&0:, 34:&1.5&ISP&2\nl
1981B&NGC 4536&b&n&56&0.41$\pm$0.14&Undetermined&3\nl
1983G&NGC 4753&s&n:&-2:&2.0&Upper limit&4\nl
1986G&NGC 5128&b&s&-9,-8&5.2&ISP&5\nl
1992A&NGC 1308&s&n&15:,100:&0.3$\pm$0.3&Undetermined&6\nl
1994D&NGC 4526&b&n&-10&0.35$\pm$0.2&Undetermined&7\nl
1994ae&NGC 3370&b&n:&$\geq 30$&0.3&Upper limit&7\nl
1995D&NGC 2962&b&n:&14, 41&0.2&Upper limit&7\nl
1996X&NGC 5061&s&n&-7,30&0.3$\pm$0.3&Maybe?&8\nl
1997dt&NGC 7448&s&n:&$\geq 0$:&?&Yes?&9\nl
1998bu&M 96&b&n&-4,-3,-2&2.1$\pm$0.1&ISP&10\nl
1999by&NGC 2841&s&s&-2,-1,0&0.8$\pm$0.1&Yes&11\nl
\enddata
\tablenotetext{a}{Spec -- Type of measurement, broadband (b) or spectropolarimetric (s).}
\tablenotetext{b}{Sub -- Strength of SN -- normal(n) or subluminous (s)}
\tablenotetext{c}{Approximate number of days past maximum light observations were taken.}
\tablenotetext{d}{Detection -- Whether or not this is a detection of 
polarization intrinsic to the SN.}  
\tablenotetext{}{References--(1) Wolstencroft \& Kemp 1972; (2) Shakhovskoi 1976; (3) Shapiro \& Sutherland 1982; (4) McCall \etal\ 1984a; (5) Hough \etal\ 1987; (6) Spyromilio \& Bailey 1993; (7) Wang \etal\ 1996; (8) Wang \etal\ 1997b (9) Leonard \etal\ 2000a; (10) Hernandez \etal\ 2000; (11) this work}
\end{deluxetable}

\begin{deluxetable}{lrrrrrrrrr}
\tablewidth{0pt}
\tablecaption{Stars within a $5\deg$ radius of SN 1999by\label{polstars-table}}
\tablehead{
\colhead{Name\tablenotemark{a}}& 
\colhead{Glon\tablenotemark{b}}&
\colhead{Glat\tablenotemark{c}}& 
\colhead{pol\tablenotemark{d}}&
\colhead{GPA\tablenotemark{e}}&
\colhead{EPA\tablenotemark{f}}& 
\colhead{$V$\tablenotemark{g}}&
\colhead{SpT\tablenotemark{h}}&
\colhead{$A_V$\tablenotemark{i}}&
\colhead{$\mu$\tablenotemark{j}}\\
\colhead{}& 
\colhead{$\deg$}&
\colhead{$\deg$}& 
\colhead{\%}&
\colhead{$\deg$}&
\colhead{$\deg$}& 
\colhead{mag}&
\colhead{}&
\colhead{mag}&
\colhead{mag}
}
\startdata
82621&164.82&45.87&0.05&81.3&6.0&4.6&A2V &0.0&3.0\nl
82328&165.42&45.68&0.01&11.1&115.0&3.2&F6IV&0.0&1.4\nl
77770&169.27&41.90&0.37&79.0&175.0&7.5&B2IV&0.0&10.6\nl
SN 1999by&166.91&44.12&&&&&&&\nl
\enddata
\tablenotetext{a}{Name --- HD Name}
\tablenotetext{b}{Glon --- Galactic longitude}
\tablenotetext{c}{Glat --- Galactic latitude}              
\tablenotetext{d}{pol --- Degree of polarization}           
\tablenotetext{e}{GPA --- Position angle of the E vector in galactic coordinates}  
\tablenotetext{f}{EPA  --- Position angle of the E vector in equatorial coordinates}      
\tablenotetext{g}{Vm --- Visual magnitude}   
\tablenotetext{h}{SpT --- Spectral type}       
\tablenotetext{i}{Av --- Visual absorption}         
\tablenotetext{j}{$\mu$ --- Distance modulus (m --- M = 5 log(r/10))}         
\tablenotetext{}{Data are from Mathewson \& Ford (1978) at http://tarantella.gsfc.nasa.gov/cgi-bin/viewer/specify.pl?file=catalog.dat\&catalog=2034A}
\end{deluxetable}


\newpage
\clearpage

\begin{thebibliography}{99}
\bibitem{} Arbour, R., Papenkova, M., Li, W. D., Filippenko, A. V., \&
Armstrong, M. 1999, IAUC 7156
\bibitem{} Benz W., Cameron, A. G. W., Press, W. H., \& Bowers, R. L. 1990, ApJ, 348, 647
\bibitem{} Bonanos, A., Garnavich, P., Schlegel, E., Jha, S., Challis, P., 
Kirshner, R., Hatano, K., \& Branch, D. 1999, AAS, 195, 3806
\bibitem{} Branch D., \& Tammann G.A, 1992, ARAA, 30, 359
\bibitem{} Burstein, D., \& Heiles, C. 1982, AJ, 87, 1165 
\bibitem{} Chandrasekhar, S. 1960, Radiative Transfer, (New York: Dover) 
\bibitem{} Collela, P., \& Woodward, P.R. 1984, J.Comp.Phys., 54, 174 
\bibitem{} Cropper, M., \etal\ 1988, MNRAS, 231, 695 
\bibitem{} Dominguez I., \& H\"oflich P. 2000, ApJ, 528, 854
\bibitem{} Filippenko, A. V., \etal\  1992, AJ, 104, 1543 (F92)
\bibitem{} Gerardy, C.L, H\"oflich P., Fesen R., \&  Wheeler J.C. 2000, ApJ, in preparation
\bibitem{} Goodrich, R. G. 1991, PASP, 103, 1314
\bibitem{} Hamuy, M., \etal\ 1994, AJ, 108, 2226
\bibitem{} Hamuy, M., \etal\ 1996a, AJ, 112, 2391
\bibitem{} Hamuy, M., Phillips, M. M., Maza, J., Suntzeff, N. B., 
Schommer, R. A., \& Avil\'es, R. 1995, AJ, 109, 1
\bibitem{} Hamuy, M., Phillips, M. M., Schommer, R. A., Suntzeff, N. B., 
Maza, J., \& Avil\'es, R. 1996b, AJ, 112, 2398
\bibitem{} Hernandez, I., \etal\ 2000, MNRAS, submitted (astro-ph/0007022)
\bibitem{} \hof , P. 1991, A\&A, 246, 481 (H91)
\bibitem{} H\"{o}flich, P. 1995a, ApJ, 440, 821
\bibitem{} H\"{o}flich, P. 1995b, ApJ, 443, 89
\bibitem{} H\"oflich, P., Dominik, C., Khokhlov, A., M\"uller, E., 
\& Wheeler, J.C. 1996, $17^{th}$ {\it Texas Symposium on Relativistic 
Astrophysics},  Annals of the New York Academy of Science, 759, 348
\bibitem{} H\"{o}flich, P., \& Khokhlov, A. 1996, ApJ, 457, 500  (HK96)
\bibitem{} H\"{o}flich, P., Khokhlov, A., \& Wheeler, J. C. 
1995a, ApJ, 444, 831 (HKW95)
\bibitem{}  H\"{o}flich, P., Khokhlov A., Wheeler, J. C., Phillips, M. M.,
Suntzeff, N. B., \& Hamuy, M. 1996, ApJ, 472, L81
\bibitem{} \hof , P., M\"uller, E., \& Khokhlov, A. M. 1993, 
A\&A, 268, 570 (HMK93)
\bibitem{} H\"oflich, P., Nomoto, K., Umeda, H., \& Wheeler, J.C. 2000, 
ApJ, 528, 854
\bibitem{} H\"oflich, P., Wang, L., Wheeler, J. C. 1999, ApJ, 521, 179
\bibitem{} H\"oflich, P., Wheeler, J. C., Hines, D., \& Trammell, S. 1995b, ApJ, 459, 307
\bibitem{} Howell, D. A. 2000a, PhD Thesis, University of Texas
\bibitem{} Howell, D. A. 2000b, in preparation
\bibitem{} Hough, J. H., Bailey, J. A., Rouse, M. F., \& Whittet, D. C. B. 1987, MNRAS, 227, 1P
\bibitem{} Hoyle, F., \& Fowler, W. A. 1960, ApJ, 132, 565
\bibitem{} Iben, I., \& Tutukov, A. V. 1984, ApJS, 54, 335
\bibitem{} Iben, I., Tutukov, A. V., \& Fedorova, A. 1998, ApJ, 503, 344 
\bibitem{} Iwamoto, K., Brachwitz, F., Nomoto, K., Kishimoto, N., Umeda, H., 
Hix, W. R., \& Thielemann, F. K. 1999, ApJS 125, 439
\bibitem{} Jeffery, D. 1991, ApJ, 375, 264  (J91)
\bibitem{} Karp, A. H., Lasher, G., Chan, K. L., \& Salpeter, E. E. 1977, ApJ, 214, 161
\bibitem{}  Khokhlov, A. 1991, A\&A, 245, 114
\bibitem{}  Khokhlov, A., M\"uller, E., \& \hof , P. 1993, A\&A, 270, 223
\bibitem{}  Khokhlov, A. 1995, ApJ, 449, 695 
\bibitem{}  Khokhlov, A. 2000, ApJ, submitted (astro-ph/0008463)
\bibitem{}  Kurucz, R.L. 1995, {CD-23}, Center for Astrophysics
\bibitem{} Kawai, Y., Saio, H., \& Nomoto, K. 1987, ApJ, 315, 229 
\bibitem{} Leibundgut, B., \etal\ 1993, AJ, 105, 301
\bibitem{}  Lentz, E. J., Baron, E., Branch, D., Hauschildt, P., \& 
Nugent, P. 1999, ApJ, 527, 746
\bibitem{} Leonard, D.C., Filippenko, A. V., \& Matheson, T. 2000a in 
{\it Cosmic Explosions}, eds. S. S. Holt \& W. W. Zhang, 
(Melville, NY: AIP), p. 165
\bibitem{} Leonard, D.C., Filippenko, A. V., Ardila, D. R., \& Brotherton, M. S. 2000b, ApJ, submitted, astro-ph/0009285  
\bibitem{} Li, W.-D., Filippenko, A. V., Treffers, R. R., Riess, A. G., 
Hu, J., \& Qiu, Y.  2000, ApJ, submitted (astro-ph/0006292)
\bibitem{} Li, W.-D., Modjaz, M., King, J. Y., Papenkova, M., Johnson, R. A., 
Friedman, A., Treffers, R. R., \& Filippenko, A. V. 1999, IAUC 7126
\bibitem{} Livne, E. 1999, ApJ, 527, L97 
\bibitem{} Lucy, L.B., \& Solomon, P.M. 1970, ApJ  159, 879
\bibitem{} Mahaffy, J. H., \& Hansen, C. J. 1975, ApJ, 201, 695
\bibitem{} Marietta, E., Burrows, A., \& Fryxell, B. 2000, ApJS, 128, 615
\bibitem{} Mathewson, D. S., \& Ford, V. 1970, Mem. RAS, 74, 139 
\bibitem{} Mathewson, D. S., Ford, V., Klare G., Neckel, T., \& Krautter, J.  1978, Bull. Inform. CDS, 14, 115 
\bibitem{} Mazzali, P. A., \etal\ 1997, MNRAS, 284, 151
\bibitem{} McCall, M. L., Reid, N., Bessell, M. S., \&
Wickramasinghe, D. 1984a, MNRAS, 210, 839
\bibitem{} McCall, M. L., Reid, N., Bessell, M. S., \& 
Wickramasinghe, D. 1984b, MNRAS, 211, 991
\bibitem{} Mihalas, D. 1978, {\it Stellar Atmospheres}, (San Francisco: 
Freeman)
\bibitem{} Miller, J. S., Robinson, L. B., \& Goodrich, R. W. 1988, in 
{\it Instrumentation for Ground-Based Optical Astronomy, Present and Future,} 
ed. L.B. Robinson, (New York: Springer-Verlag)
\bibitem{} Milne P., The, L.S., \& Leising, M.D. 1999, ApJS 124, 503
\bibitem{} Mochkovitch, R. Guerrero, J., \& Segretain, L. 1997,  in {\it Thermonuclear Supernovae}, eds. Ruiz-Lapuente, Canal, \& Isern 1997, (Dordrecht: Kluwer), pp. 187  
\bibitem{} Mochokovitch, R., \& Livio, M. 1990, A\&A, 236, 378
\bibitem{} Modjaz, M., \etal\ 2000, PASP, submitted (astro-ph/0008012)
\bibitem{} M\"uller, E., \& Eriguchi, Y. 1985, A\&A, 152, 325
\bibitem{} Niemeyer, J. C., \& Hillebrandt, W. 1995, ApJ, 452, 779
\bibitem{} Nomoto, K. 1980, ApJ, 248, 798
\bibitem{} Nomoto, K., \& Iben, I. J. 1985, ApJ, 297, 531
\bibitem{} Nomoto, K., \& Kondo Y. 1991, ApJ, 367, L19
\bibitem{} Nomoto, K., \& Sugimoto, D. 1977, PASJ, 29, 765
\bibitem{} Nomoto, K., Thielemann, F. K., \& Yokoi, K. 1984, ApJ, 286, 644
\bibitem{} Nugent, P., Baron, E., Hauschildt, P., \& Branch., D. 1997, 
ApJ, 485, 812
\bibitem{} Nugent, P., Phillips, M. M., Baron, E., Branch, D., 
\& Hauschildt, P.  1995, ApJ, 455L, 147 (N95)
\bibitem{} Paczy\'nski  B. 1985, in {\it Cataclysmic Variables and Low-Mass X-Ray Binaries}, eds. D.Q. Lamb \& J. Patterson,  (Dordrecht: Reidel) p. 1
\bibitem{} Perlmutter, S., et al.\ 1997, ApJ, 483, 565
\bibitem{} Perlmutter, S., et al.\ 1999, ApJ, 517, 565 
\bibitem{} Phillips, M. 1993, ApJ, 413, L105  
\bibitem{} Rasio, F. A., \& Shapiro, S. L. 1995, ApJ, 438, 887
\bibitem{} Riess, A. G., et al.\ 1999, AJ, 117, 707
\bibitem{} Riess, A. G., Press, W. H., \& Kirshner, R. P. 1996, ApJ, 473, 88
\bibitem{} Ruiz-Lapuente, P., Jeffery, D. J., Challis, P., Filippenko, A. V., 
Kirshner, R. P., Ho, L. C., Schmidt, B. P., Sanchez, F., \& Canal, R. 1993, 
Nature 365, 728
\bibitem{} Saffer, R. A., Livio, M., \& Yungelson, L. R., 1998, ApJ, 502, 394
\bibitem{} Saio, H., \& Nomoto, K. 1985, A\&A, 150, L21
\bibitem{} Saio, H., \& Nomoto, K. 1998, ApJ, 500, 388
\bibitem{} Savage, B. D., \& Mathis, J. S. 1979, ARAA, 17, 73
\bibitem{} Schaefer, B. E. 1996, ApJ, 464, 404 
\bibitem{} Schlegel, D. J., Finkbeiner, D. P., \& Davis, M. 1998, ApJ, 500, 525
\bibitem{} Schmidt, B. P., et al. 1998, ApJ 507, 46
\bibitem{} Serkowski, K., Mathewson, D., \& Ford, V. L. 1975, ApJ, 196, 261
\bibitem{} Shakhovskoi, N. M. 1976, Soviet Astron. Lett., 2, 107
\bibitem{} Shapiro, P. R., \& Sutherland, P. G. 1982, ApJ, 263, 902
\bibitem{} Simmons, J. F. L., \& Stewart, B. G. 1985, A\&A, 142, 100
\bibitem{} Sobolev, V. V. 1957, Sov. Astron. 1, 297
\bibitem{} Spyromilio, J., \& Bailey, J. 1993, PASAu, 10, 263S
\bibitem{} Thielemann F.-K., Nomoto K., \& Hashimoto M. 1996, ApJ, 460, 408
\bibitem{} Trammell, S. R. 1994, PhD Thesis, University of Texas, pp. 246-260
\bibitem{} Tran, H. D. 1995, ApJ, 440, 578
\bibitem{} Treffers, R. R., Peng, C. Y., Filippenko, A. V., \&
Richmond, M. W., 1997, IAUC 6627
\bibitem{} Turatto, M., \etal\ 1996, MNRAS, 283, 1
\bibitem{} Turatto, M., Piemonte, A., Benetti, S., Cappellaro, E., 
Mazzali, P. A., Danziger, I. J., \& Patat, F. 1998, AJ, 116, 2341
\bibitem{} Umeda, H., Nomoto, K., Yamaoka, H., \& Wanajo S. 1999,  ApJ, 513, 861 
\bibitem{}  Van de Hulst, H.C. 1957, { \it Light Scattering by Small Particles}, (New York: Wiley)
\bibitem{} Wang, L., Baade, D., Fransson, C., \hof, P., \& Wheeler, J. C. 2000b, 
ApJ, submitted 
\bibitem{} Wang, L., Howell, D. A., \hof, P., \& Wheeler, J. C. 2000a, 
ApJ, submitted (astro-ph/9912033)
\bibitem{} Wang, L., \& Wheeler, J. C. 1996, ApJ, 462, L27 
\bibitem{} Wang, L., Wheeler, J. C., \& H\"oflich, P. 1997a, 
ApJ, 476, L27 (WWH97)
\bibitem{}  Wang, L., Wheeler, J. C., \& H\"oflich, P. 1997b, in SN1987A, Ten
Years After, eds. M. M. Phillips \& N. Suntzeff, (New York: Kluwer), in press
\bibitem{} Wang, L., Wheeler, J. C., Li, Z., \& Clocchiatti, A.  1996, ApJ, 467, 435
\bibitem{} Webb, W., Malkan, M., Schmidt, G., \& Impey, C. 1993, ApJ, 419, 494
\bibitem{} Webbink, R. F. 1984, ApJ, 277, 355  
\bibitem{} Wheeler, J. C., \hof , P., Harkness, R. P., \& 
Spyromilio, J. 1998, ApJ, 496, 908
\bibitem{} White, R. L. 1979, ApJ, 229, 954
\bibitem{} Whittet, D. C. B., Martin, P. G., Hough, J. H., Rouse, M. F., Bailey, J. A., \& Axon, D. J. 1992, ApJ, 386, 562
\bibitem{} Wolstencroft, R. D., \& Kemp, J. C. 1972, Nature, 238, 452
\bibitem{} Woosley, S. E., \& Weaver, T. A. 1994, ApJ, 423, 371
\bibitem{} Woosley  S.E., \& Weaver  T.A. 1986, ARAA, 24, 205
\bibitem{} Yamaoka H., Nomoto K., Shigeyama T., \& Thielemann F., 1992, ApJ, 393, 55    

\end{thebibliography}
\end{document}